# Magnetization Dynamics, Throughput and Energy Dissipation in a Universal Multiferroic Nanomagnetic Logic Gate with Fan-in and Fan-out

Mohammad Salehi Fashami[1], Jayasimha Atulasimha[1*] and Supriyo Bandyopadhyay[2]

[1]Department of Mechanical and Nuclear Engineering, [2]Department of Electrical and Computer Engineering, Virginia Commonwealth University, Richmond VA 23284, USA.

**Abstract**

The switching dynamics of a multiferroic nanomagnetic NAND gate with fan-in/fan-out is simulated by solving the Landau-Lifshitz-Gilbert (LLG) equation while neglecting thermal fluctuation effects. The gate and logic wires are implemented with dipole-coupled 2-phase (magnetostrictive/piezoelectric) multiferroic elements that are clocked with electrostatic potentials of ~50 mV applied to the piezoelectric layer generating 10 MPa stress in the magnetostrictive layers for switching. We show that a pipeline bit throughput rate of ~ 0.5 GHz is achievable with proper magnet layout and sinusoidal four-phase clocking. The gate operation is completed in 2 ns with a latency of 4 ns. The total (internal + external) energy dissipated for a single gate operation at this throughput rate is found to be only ~ 1000 kT in the gate and ~3000 kT in the 12-magnet array comprising two input and two output wires for fan-in and fan-out. This makes it respectively 3 and 5 orders of magnitude more energy-efficient than complementary-metal-oxide-semiconductor-transistor (CMOS) based and spin-transfer-torque-driven nanomagnet based NAND gates. Finally, we show that the dissipation in the external clocking circuit can always be reduced asymptotically to zero using increasingly slow adiabatic clocking, such as by designing the RC time constant to be 3 orders of magnitude smaller than the clocking period. However, the internal dissipation in the device must remain and cannot be eliminated if we want to perform fault-tolerant classical computing.

--------------------------------------------------------------------------------

[*] Corresponding author. E-mail: jatulasimha@vcu.edu

## I. Introduction

A major challenge in designing digital computing machinery is to reduce energy dissipation during the execution of a computational step since excessive dissipation is the primary impediment to device downscaling envisioned in Moore's law. A state-of-the-art CMOS nanotransistor presently dissipates over 50,000 kT ($2\times10^{-16}$ Joules) of energy at room temperature to switch in *isolation*, and over $10^6$ kT ($4.2\times10^{-15}$ Joules) to switch in a circuit at a clock rate of few GHz [1], which makes further downscaling problematic. Therefore, nanomagnet-based computing and signal processing are attracting increasing attention. Single-domain nanomagnets are intrinsically more energy-efficient than transistors as logic switches and do not suffer from current leakage that results in *standby power dissipation.* Consequently, magnetic architectures hold the promise of outpacing transistor-centric architectures in energy-efficient computing.

A nanomagnetic logic switch is typically implemented with a shape-anisotropic nanomagnet that has two stable magnetization orientations along the easy axis. These two magnetizations encode the logic bits 0 and 1. The nanomagnet's advantage over the transistor accrues from the fact that when the magnetization is flipped, all the spins (information carriers) in a single-domain nanomagnet rotate in *unison* like a giant classical spin (because of exchange interaction between spins), so that ideally the magnet has but a *single* degree of freedom [2]. Accordingly, the minimum energy dissipated in switching it non-adiabatically is ~ *kTln(1/p)* where *p* is the static error probability associated with random switching due to thermal fluctuations. In contrast, all the different electrons or holes (information carriers) in a transistor act independently, so that the *minimum* energy dissipated in switching a transistor non-adiabatically is *NkTln(1/p)* [2], where *N* is the number of information carriers (electrons or holes) in the transistor. The advantage of the single-domain nanomagnet thus accrues not from any *innate* advantage of spin over charge as an information carrier, but from the fact that spins mutually interact in a way which reduces the degrees of freedom, and hence the energy dissipation.

There are numerous ways of implementing nanomagnetic logic (NML) [3], [4]. In one type of architecture, termed 'magnetic quantum cellular automata' [5], logic gates are configured by placing nanomagnets in specific geometric patterns on a surface so that dipole interactions between neighbors elicit the desired logic operations on the bits encoded in the magnetization orientations of the nanomagnets[4] , [5]. This approach is the same as that envisioned in the Single Spin Logic (SSL) paradigm, where exchange interaction between spins played the role of dipole interaction between magnets, while up- and down-spin polarizations encoded logic bits [6]. However, SSL required cryogenic operation while NML can operate at room temperature. Curiously, the *minimum* energy dissipated per bit flip in NML is the same as that in SSL because a single spin and a giant classical spin have the same degree of freedom. This is a remarkable feature that makes NML particularly attractive. Of course, the actual energy dissipated per bit flip in NML will always be somewhat higher than *kTln(1/p)* because of internal dynamics such as spin-orbit interaction giving rise to Gilbert damping within the magnet, but that is still a small price to pay for room temperature operation.

### a. Magnet switching schemes

Unfortunately, the nanomagnet's advantage over the transistor will be squandered if the method employed to switch the nanomagnet becomes so energy-inefficient that the energy dissipated in the switching circuit vastly exceeds the energy dissipated in the nanomagnet. In the end, this can make magnetic architectures less energy-efficient than transistor based architectures, thereby defeating the entire purpose of using magnetic switches. Therefore, the switching scheme is vital.

Magnets are typically switched with either a magnetic field generated by a current [7], or with spin transfer torque [8], or with domain wall motion induced by a spin polarized current [9]. In the first approach, a local magnetic field is generated by a local current based on Ampere's law:

$$I = \int_c \vec{H} \cdot d\vec{l} \qquad (1)$$

where the integral is taken around the current loop.

Now, the minimum magnetic field $\vec{H}_{min}$ required to flip a magnet can be estimated by equating the magnetic energy in the field to the energy barrier $E_b$ separating the two magnetization directions encoding the bits 0 and 1, i.e.

$$\mu_0 M_s H_{min} \Omega = E_b \quad (2).$$

Here $\mu_0$ is the permeability of free space, $M_s$ is the saturation magnetization which we assume is $10^5$ A/m (typical value for nickel or cobalt), and $\Omega$ is the nanomagnet's volume which we assume is 100 nm × 100 nm × 10 nm since such a volume results in a single domain ferromagnet at room temperature. The energy barrier $E_b$ is determined by the static bit error probability $e^{-E_b/kT}$ that we can tolerate. For reasonable error probability (< $e^{-30}$), we will need that $E_b \geq 30kT$. This yields from Equations (1) and (2) that $I_{min}$ = 6 mA if we assume that the loop radius is 100 nm so that it can comfortably encircle the magnet. The resistance of loop will be ~50 ohms if we assume that it is made of silver (lowest resistivity among metals; $\rho = 2.6 \mu$S-cm) and has a wire radius of 10nm, so that the energy dissipated – assuming that the magnet flips in 1 ns – is at least 1.8 pJ, or ~ $4 \times 10^8$ kT, which makes the magnet switch ~400 times more dissipative than a state-of-the-art transistor switch because of the inefficient switching scheme. There is also another disadvantage; the magnetic field cannot be confined to small spaces, which means that individual magnets cannot be addressed unless the magnet density is sparse (magnet separation ≥ 0.5 μm). That not only reduces device density, but might make dipole interaction between magnets so weak as to make magnetic quantum cellular automata inoperable. Therefore, this method is best adapted to addressing not individual magnets, but groups of (closely spaced) magnets together, as envisioned in ref. [7]. However, that approach makes magnetic quantum cellular automata architecture *non-pipelined* and hence very slow [10]. In the end, this is clearly a sub-optimal method of switching magnetic switches.

Spin transfer torque (STT) is better adapted to addressing individual magnets since it switches magnets with a spin polarized current passed directly through the magnet. It dissipates about $10^8$ kT of energy to switch a single-domain nanomagnet in ~ 1 ns, even when the energy barrier within the magnet is only ~ 30 kT [11]. Thus, it is hardly better than the first approach in terms of energy efficiency. A more efficient method of switching a magnet is by inducing domain wall motion by passing a spin polarized current through the magnet. There is at least one report of switching a multi-domain nanomagnet in 2 ns by this approach while dissipating $10^4$ kT – $10^5$ kT of energy [12]. This makes it 1-2 orders of magnitude more energy-efficient than a transistor in a circuit.

Recently, we devised a much more efficient magnet switching scheme. We showed that a 2-phase multiferroic nanomagnet, consisting of a piezoelectric layer elastically coupled with a magnetostrictive layer, can be switched by applying a small voltage of few mV to the piezoelectric layer [13], [14]. This voltage generates uniaxial strain in the piezoelectric layer that is transferred almost entirely to the magnetostrictive layer by elastic coupling if the latter layer is much thinner than the former. Uniaxiality can be enforced in two ways: either by applying the electric field in the direction of expansion and contraction ($d_{33}$ coupling) or by mechanically clamping the multiferroic in one direction and allowing expansion/contraction in the perpendicular direction through $d_{31}$ coupling when the voltage is applied *across* the piezoelectric layer. The substrate is assumed to be a soft material (e.g. a polymer) that allows uniaxial expansion/contraction. The uniaxial strain/stress will cause the magnetization of the magnetostrictive layer to rotate by a large angle. Such rotations can be used for Bennett clocking of NML gates for logic bit propagation [13]. In ref. [14], [15], we showed that the energy dissipated in the magnet and clock together is a few hundreds of kT for a switching delay of 1 ns or less. This makes it one of the most energy-efficient magnet switching schemes.

In this paper, we have studied the switching dynamics of a NAND gate with fan-in/fan-out wires implemented with multiferroic elements and calculated the energy dissipation in the entire block assuming low enough temperature when effects of thermal fluctuations can be neglected. At room temperature, thermal fluctuations will act as a random magnetic field that will increase the switching error probability and mandate higher stress levels (along with larger energy dissipation) for reliable gate operation. This study is deferred to a future date.

The present paper is organized as follows: In Section II, the theoretical framework for studying magnetization dynamics in the NAND gate with fan-in and fan-out wires is discussed. Section III presents and discusses simulation results. Section IV discusses energy dissipation and strategies to reduce the energy overhead in the clocking circuit. Finally in section V, we present our conclusions.

## II. Magnetization Dynamics in a Multiferroic Nanomagnetic Logic Gate with Fan-in and Fan-out

Consider an all-multiferroic NAND gate with fan-in and fan-out wires as shown in Fig 1. Each multiferroic element has the shape of an elliptical cylinder with homogeneous magnetization $\vec{M}(\vec{r})$ in the magnetostrictive layer. We assume that the piezoelectric layer has a thickness of 40 nm and the magnetostrictive layer has a thickness of 10 nm, which will ensure that strain generated in the piezoelectric layer is mostly transferred to the magnetostrictive layer through elastic coupling. There are mechanical clamps along the minor axis of the magnet (not shown) that prevent expansion/contraction along that sideward direction, so that application of a voltage across the piezoelectric layer generates uniaxial stress along the major axis via the $d_{31}$ coupling. The magnets are fabricated on a soft substrate that does not hinder expansion and contraction of the elements along any direction by clamping from below.

If the planar dimensions of each element are ~ 101.75 nm × 98.25 nm, the exchange coupling penalty precludes the formation of multi-domain states in the magnetostrictive layer [16], [17] so that we can model it as a single-domain nanomagnet. The shape anisotropy of the element gives rise to an energy barrier of 32 kT (at room temperature) between the two stable orientations along the easy axis (major axis) of the ellipse.

The magnetization dynamics of any nanomagnet under the influence of an effective field $\vec{H}_{eff}$ acting on it is described by the Landau-Lifshitz -Gilbert (LLG) equation [17]:

$$\frac{d\vec{M}(t)}{dt} = -\nu \vec{M}(t) \times \vec{H}_{eff}(t) - \frac{\alpha \nu}{M_s}\left[\vec{M}(t) \times \left(\vec{M}(t) \times \vec{H}_{eff}(t)\right)\right] \quad (3).$$

Here $\vec{H}_{eff}^{i}$ is the effective magnetic field on the i$^{th}$ nanomagnet, which is the partial derivative of its total potential energy ($U_i$) with respect to its magnetization ($\vec{M}_i$), $\nu$ is the gyromagnetic ratio, $M_s$ is the saturation magnetization of the magnetostrictive layer and $\alpha$ is the Gilbert damping factor [18] associated with internal dissipation in the magnet when its magnetization rotates. Accordingly,

$$\vec{H}_{eff}^{i}(t) = -\frac{1}{\mu_0 \Omega}\frac{\partial U_i(t)}{\partial \vec{M}_i(t)} = -\frac{1}{\mu_0 M_s \Omega}\nabla_{\vec{m}} U_i(t) \quad , \quad (4).$$

where $\Omega$ is the volume of any nanomagnet (only that of the magnetostrictive layer) in the chain shown in Fig 1. The total potential energy of any element in this chain is given by:

$$U_i(t) = \underbrace{\sum_{j \neq i} E_{dipole-dipole}^{i-j}(t)}_{E_{dipole}} + \underbrace{\left(\frac{\mu_0}{2}\right)\left[M_s^2 \Omega\right]\left(N_{d\_xx}\left[\sin\theta_i(t)\cos\phi_i(t)\right]^2 + N_{d\_yy}\left[\sin\theta_i(t)\sin\phi(t)_i\right]^2 + N_{d\_zz}\left[\cos\theta_i(t)\right]^2\right)}_{E_{shape-anisotropy}}$$
$$\underbrace{-\left(\frac{3}{2}\lambda_s \sigma_i \Omega\right)\sin^2\theta_i(t)\sin^2\phi_i(t)}_{E_{stress-anisotropy}} + \text{other terms} \quad (5).$$

Here $E_{dipole-dipole}^{i-j}$ is the dipole-dipole interaction energy due to interaction between the i-th and j-th magnets, $E_{shape-anisotropy}$ is the shape anisotropy energy due to the elliptical shape of the multiferroic element, and $E_{stress-anisotropy}$ is the stress anisotropy energy caused by the stress $\sigma$ transferred to the magnetostrictive layer of the multiferroic upon application of an electrostatic potential to the piezoelectric layer. The quantities $N_{d-kk}$ are the demagnetization factors in the k-th direction. We assume that the magnetostrictive layer is polycrystalline so that we can neglect magnetocrystalline anisotropy. The quantity $\lambda_s$ is the magnetostrictive coefficient and the quantities $\theta$ and $\phi$ are the polar and azimuthal angles of the magnetization vector $\vec{M}_i$. We assume that the major axis (easy axis) of the ellipse is in the y-direction, the minor axis (in-plane hard axis) is in the x-direction and the out-of-plane hard axis is in the z-direction as shown in Fig. 1(b). The 'other terms' in the above equation are time-dependent (they include the chemical potential, the mechanical potential due to stress/strain etc.), but do not depend on $\vec{M}_i$ (or $\theta_i, \phi_i$) and hence do not affect $\vec{H}_{eff}^i$.

Consider two adjacent multiferroic elements in the chain (labeled as the $i^{th}$ and $j^{th}$ element), whose magnetizations subtend an angle of $\theta_i, \phi_i$ and $\theta_j, \phi_j$ respectively with the positive z-direction and x-direction in the x-y plane. The dipole-dipole interaction energy is [19]:

$$E_{dipole-dipole}^{i-j}(t) = \frac{\mu_0 M_s^2 \Omega^2}{4\pi |\vec{r}_{i-j}|^3} \left[ \left(\vec{m}_i(t).\vec{m}_j(t)\right) - \frac{3}{|\vec{r}_{i-j}|^2}\left(\vec{m}_i(t).\vec{r}_{i-j}\right)\left(\vec{m}_j(t).\vec{r}_{i-j}\right) \right] \qquad (6).$$

where $\vec{r}_{i-j}$ is the vector distance between the $i^{th}$ and $j^{th}$ magnet and $\vec{m}_k$ is the magnetization of the k-th magnet normalized to $M_s$.

For two neighboring magnets whose in-plane hard axes are collinear with the line joining their centers, the dipole coupling energy is:

$$E_{dipole-dipole}^{i-j}(t) = \frac{\mu_0 M_s^2 \Omega^2}{4\pi r^3} \sum_j \begin{bmatrix} -2\left(\sin\theta_i(t)\cos\phi_i(t)\right)\left(\sin\theta_j(t)\cos\phi_j(t)\right) \\ +\left(\sin\theta(t)_i \sin\phi_i(t)\right)\left(\sin\theta_j(t)\sin\phi_j(t)\right) \\ +\cos\theta_i(t)\cos\theta_j(t) \end{bmatrix} \qquad (7).$$

If the line joining the centers subtends an angle $\gamma$ with their hard axes as shown in Fig 1 (b), the dipole coupling energy is:

$$E_{dipole-dipole}^{i-j}(t) = \frac{\mu_0 M_s^2 \Omega^2}{4\pi r^3} \sum_j \begin{bmatrix} (\sin\theta_i(t)\cos\phi_i(t))(\sin\theta_j(t)\cos\phi_j(t))\left(-2(\cos\gamma)^2+(\sin\gamma)^2\right) \\ +(\sin\theta_i(t)\sin\phi_i(t))(\sin\theta_j(t)\sin\phi_j(t))\left(-2(\sin\gamma)^2+(\cos\gamma)^2\right) \\ +\{(\sin\theta_i(t)\cos\phi_i(t))(\sin\theta_j(t)\sin\phi_j(t))+(\sin\theta_j(t)\cos\phi_j(t))(\sin\theta_i(t)\sin\phi_i(t))\} \\ \times(-3\sin\gamma\cos\gamma)+\cos\theta_i(t)\cos\theta_j(t) \end{bmatrix} \qquad (8).$$

where $r$ is the separation between their centers.

The total potential energy of a magnet given in Equation (5) is used to find the effective field $\vec{H}_{eff}$ acting on it in accordance with Equation (4):

$$H^i_{eff-x}(t) = \sum_{j \neq i} -\frac{1}{\mu_0 M_s \Omega} \frac{\partial E^{i-j}_{dipole-dipole}(t)}{\partial m_x(t)} - M_s(N_{d-xx})\sin\theta_i(t)\cos\phi_i(t)$$

$$H^i_{eff-y}(t) = \sum_{j \neq i} -\frac{1}{\mu_0 M_s \Omega} \frac{\partial E^{i-j}_{dipole-dipole}(t)}{\partial m_y(t)} - M_s(N_{d-yy})\sin\theta_i(t)\sin\phi_i(t) + \left(\frac{3}{\mu_0 M_s}\lambda_s\right)\sigma_i(t)\sin\theta_i(t)\sin\phi_i(t) + H_{bias} \quad (9).$$

$$H^i_{eff-z}(t) = \sum_{j \neq i} -\frac{1}{\mu_0 M_s \Omega} \frac{\partial E^{i-j}_{dipole-dipole}(t)}{\partial m_z(t)} - M_s(N_{d-zz})\cos\theta_i(t)$$

where $H_{bias}$ is any external time-invariant magnetic field applied in the y-direction (see later discussion as to when and why it might be needed).

Further, it appears from equation (9) that a stress applied in the y-direction, produces only a $H_{eff}$ along the y-direction and could not rotate the magnetization along the x or z-direction. However, a deeper analysis of its effect on the magnetization shows this is not true. We will explain this by writing the stress anisotropy energy term and the associated $H_{eff}$ in Cartesian co-ordinates as:

$$E_{stress-anisotropy} = -\left(\frac{3}{2}\lambda_s \sigma_i(t)\Omega\right)m_y^2(t) \qquad H_{eff\_y} = \frac{3}{\mu_0 M_s}(\lambda_s \sigma_i(t))m_y(t)$$

Now suppose a compressive stress (negative σ) is applied to a material with positive magnetostriction (λ) and the magnetization is initially directed close but not parallel to +y-axis so that $m_y \sim 1$. The $H_{eff}$ is along the -y axis and would make the magnetization rotate away from +y towards the x-direction (also lifting the magnetization vector a little out of plane; the lifting is negligible when $N_{d\_ZZ}$ is very large). However, when the magnetization rotates close to the hard (x-axis), $m_y \sim 0$, and the $H_{eff}$ due to stress vanishes. If the magnetization had initially pointed along the -y-axis, $H_{eff}$ would have been directed along the +y-axis, which would have again rotated the magnetization away from the -y-axis towards the in plane hard axis (x-axis).

The effective field $H_{eff}$ for each nanomagnet described by Equation (9) includes the effect of dipole coupling with neighboring nanomagnets, shape anisotropy and stress anisotropy. For dipole coupling, the summation is performed over nearest, second-nearest and sometimes third-nearest neighbors since the interaction with remote neighbors may not be negligible compared to that with the nearest neighbor. We elucidate this below:

In Fig 1 (a), consider a magnet marked "I". The nearest neighbor marked "II" is at a distance $r$, but the second nearest neighbor marked 'III' is at a distance of $\sqrt{5/2}r$ which results in a dipole field (proportional to $1/r^3$) that is ~25 % of the dipole field due to the magnet marked 'II'. Hence both have to be considered. Similarly, for a magnet marked "III", interactions with magnets marked "I", "II" and "IV", whose centers are respectively at distances of $\sqrt{5/2}r$, $\sqrt{3/2}r$ and $r$, are considered. For magnets marked "IV", "V", and "VII" both nearest and second nearest neighbor dipole coupling terms were considered. However, for the magnet marked "VI" only the nearest neighbors' dipole coupling was considered since the second nearest neighbor is at a distance $2r$ away and contributes only 12.5% of the interaction caused by the nearest neighbor.

The effective magnetic field $H_{eff}$ evaluated from Equation (9) is used in the scalar version of equation (3) which is described in ref. [15] to determine the magnetization state of any magnet at any instant of time. Equation (3) leads to two coupled ordinary differential equations (ODEs) for each nanomagnet. Thus, for 12 nanomagnets, 24 coupled ODEs have to be solved simultaneously. This allows us to compute the temporal evolution of the magnetization orientations $\left(\theta_i(t), \phi_i(t)\right)$ of all 12 magnets in Fig 1.

In order to flip the magnetization of any magnet, it is first subjected to stress with a voltage $V$ applied across the piezoelectric layer. The sign of the stress (compressive or tensile) depends on the sign of the magnetostrictive coefficient $\lambda_s$ of the material. For a material with positive $\lambda_s$, such as Terfenol-D, we apply a compressive (negative) stress to rotate the magnetization away from the easy axis (y-axis) towards the in-plane hard axis (x-axis). If the stress is of sufficient magnitude to overcome the shape anisotropy energy barrier, the magnetization will ultimately align along the in-plane hard axis (x-axis). Once that happens, the stress is reversed to tensile so that the magnetization relaxes back to the easy axis (y-axis), but in a direction anti-parallel to the original orientation along the easy axis. What ensures the anti-parallel orientation is that stress not only causes in-plane rotation of the magnetization vector, but also lifts it slightly out of plane in such a way that the resulting y-component of $\vec{H}_{eff}$ prefers the anti-parallel orientation over the parallel orientation. This results in a magnetization flip or bit flip.

The time evolution of the magnetization during this process is tracked by solving the LLG equation (Equation (3)) starting with the initial condition. If we choose the initial orientation of the magnetization to be *exactly* along the easy axis $(\theta = \phi = 90^0)$, then the effective field on the magnet due to stress vanishes [see Equation (9)] and therefore stress becomes ineffective in causing any rotation. Hence, we always assume that the initial state of any magnetization is never exactly along an easy axis but deflected by 0.1 degrees from the easy axis $(\theta = 90^0, \phi = \pm 89.9^0)$. Such deflections ~ 5º can easily occur because of thermal fluctuations [20] but we are conservative and assume only 0.1º deflection to ensure our simulation results are applicable over a greater confidence interval.

The energy dissipated in flipping a bit has two components:

i) Energy dissipated while applying, reversing and removing a voltage on the piezoelectric layer for generating stress. This is the energy dissipated in the clocking circuit and is given by [21]:

$$E_{clock} = \frac{1}{2}CV^2 \frac{\omega RC}{1+(\omega RC)^2} \qquad (10a).$$

where $C$ is the capacitance of the piezoelectric layer, R is the resistance of the wires and $V$ is the voltage applied across it. We assume that the voltage waveform is sinusoidal with a period $2\pi/\omega$. However, the problem with this RC circuit is that the $1/2\,(CV^2)$ energy stored in the capacitor (piezoelectric layer), when it is fully charged, minus the dissipation in the resistor, will be dissipated in the power source in each cycle. A better scheme is to use an LCR circuit where energy is merely transferred between the capacitive and inductive elements and the only energy lost per cycle is the energy dissipated in the resistive element [22]. In such a clocking circuit, the energy dissipated is:

$$E_{clock} = \pi \frac{V^2}{R} \omega (RC)^2 \tag{10b}$$

ii) Internal energy dissipated in the magnet during magnetization rotation [14, 15, 17]. This energy $E_d$ is calculated as:

$$\frac{dE_d(t)}{dt} = -\mu_0 \int_\Omega \vec{H}_{eff} \cdot \frac{d\vec{M}}{dt} \, d\Omega \tag{11}$$

By substituting Equation (3) for $\frac{d\vec{M}}{dt}$ in equation (11) and integrating one obtains:

$$E_d(\tau) = \int_0^\tau -\left(\frac{dE_d}{dt}\right) dt = \int_0^\tau \frac{\alpha \mu_0 \nu \Omega}{(1+\alpha^2)M_s} |\vec{H}_{eff}(t) \times \vec{M}(t)|^2 \, dt \tag{12}$$

This expression clearly shows that this dissipation is associated with damping in the magnet because it disappears when $\alpha = 0$.

**III. Results and Discussions**

We have used 4$^{th}$ order *Runge-Kutta* method as described in [15] to solve the system of 24 coupled ordinary differential equations for a chain of twelve dipole coupled multiferroic elements shown in Fig. 1(a). These 12 magnets comprise the NAND-gate and wiring for fan-in of 2 and fan-out of 3. The stress applied on the four nanomagnets comprising the actual gate follows a 4-phase sinusoidal clocking scheme shown in Fig 1(c). The magnets are grouped into 8 groups I through VIII. The sinusoidal clocks applied to each group and the relative phase lags between the clock signals for different groups is shown in Fig. 1(c). Clearly, a 4-phase clock is required. When the phase for the clock on magnets marked "I" goes past 90° so that the compressive stress on these magnets begins to decrease, the compressive stress on magnets marked II just begins to increase. Thus, when the stress on magnets "I" has decreased to $\sqrt{1/2}$ of the maximum applied compression, the magnets marked "II" are at a state of $\sqrt{1/2}$ of the maximum compression and have been sufficiently rotated away from the easy direction. Consequently, as compressive stress decreases to a point where the shape anisotropy begins to dominate and therefore the magnetizations of magnets marked "I" rotate towards their easy axes, their orientation is influenced strongly and ultimately uniquely determined by the orientations of the "input" magnets ensuring uni-directionality of information propagation [15].

From the time-dependent voltages on any magnet, we derive the time-dependent stresses and hence the time dependent effective fields $\vec{H}_{eff}^{i}(t)$ on each magnet. These are used to solve the LLG equation (24 coupled ODEs). The solutions yield the orientation $\theta_i(t)$, $\phi_i(t)$ of each element. The in-plane magnetization orientation ($\phi_i(t)$) of each of the 12 nanomagnets (Fig 2 a-d) is plotted to demonstrate: (i) successful NAND operation for any arbitrary input combination [(1,1), (0,0), (1,0), (0,1)] starting with the initial input state (1, 1), and (ii) the complete magnetization dynamics showing that the primitive gate operation is always completed in 2 ns and the latency is 4 ns.

In this study, we assumed that the magnetostrictive layers were made of polycrystalline Terfenol-D with material properties and dimensions given in Table I. The piezoelectric layer is assumed to be lead-zirconate-titanate (PZT) that has a reasonably large $d_{31}$ coefficient ($10^{-10}$m/V[27]), albeit also a large relative dielectric constant of 1000. Terfenol-D was chosen for its high magnetostriction [23] (even in the nanoscale [24]). The geometric parameters for the individual magnets and the array were chosen to ensure: (i) the shape anisotropy energy of the elements was sufficiently high (~0.8 eV or ~32$kT$ at room temperature) so that the bit error probability due to spontaneous magnetization flipping was very low ($\sim e^{-32} \approx 10^{-14}$), (ii) the dipole interaction energy was limited to 0.26 eV which was significantly lower than the shape anisotropy energy to prevent spontaneous flipping of magnetization, but large enough to ensure that the magnetization of the multiferroic elements always flipped to the correct orientation when stress was applied, even under the influence of random thermal fluctuations, and (iii) the maximum applied stress of 10 MPa corresponded to a stress-anisotropy energy $\frac{3}{2}[\lambda_s \sigma \Omega]$ = 172 $kT$ that was significantly larger than the shape anisotropy energy barrier of 32 $kT$. The reason why such large stress was required are: (1) some magnets (for example the magnet marked "III") had to overcome significant amount of dipole coupling from interaction with multiple neighbors to rotate close to the hard axis; (2) the stress anisotropy is least effective close to $\Phi$=0 and hence the stress had to be large to ensure fast magnetization rotation for angles close to the hard axis.

In all our simulations (Fig 2 a-d), the initial magnetizations of the nanomagnets always correspond to the ground state of the array corresponding to input bits "1" and "1". When a new input stream arrives, the input bits are changed to conform to the new inputs. Thus, at time $t$ = 0, the magnetizations of input-1 and input-2 are respectively set to (1, 1) [Fig 2(a)], (0, 0) [Fig 2 (b)], (1, 0) [Fig 2 (c)], (0, 1) [Fig 2(d)]. We then consider the time evolution of the in-plane magnetization orientations of every multiferroic nanomagnet when a 4-phase stress cycle is applied, as shown in Fig 1 (c), to clock the array. In Fig 2(a), the inputs are unchanged as input-1 = 1 and input-2 = 1. This is a trivial case as the ground state already corresponds to the correct output. But it is still important to simulate the magnetization dynamics to verify that the gate works correctly. As seen in Fig 2(a), all magnetizations rotate through ±90º to the hard axis under compressive stress and then rotate back to their initial (correct) orientations under the influence of dipole coupling as the stresses are reversed to tensile. This results in a logical NAND output of "0". As expected there is a phase (and time) lag between instants when the compressive stress reaches a maximum and the magnetization is closest to the hard axis. This is because magnetization takes a finite time to respond to the applied stress, as is evident from the LLG equations.

In Fig 2(b), the inputs are both changed so that input-1 = 0 and input-2 = 0. Therefore, all the magnets in the input wire, gate and output wire flip through 180º, rotating first through ±90º on application of a compressive stress and then further rotating through ±90º under the influence of dipole coupling. The phasing of the clock not only ensures the correct logical NAND output of "1" is reached but that the information is propagated *unidirectionally* through the input branches as well as the three output branches. The 4-phase clock achieves the following: As the compressive stress on a magnet is lowered to a point where the shape anisotropy barrier is about to be restored, the compressive stress on its right (subsequent) neighbor has already rotated it towards its hard axis. Therefore, the state of its left (previous) neighbor determines the easy direction towards which the stressed magnet will relax as the stress in lowered. This ensures unidirectional logic bit propagation as in the case of Bennett clocking [28]. Finally, Figs. 2(c) and Fig 2(d) show magnetization dynamics for the cases when one of the inputs is set to "1" while the other is set to "0". Here again the correct logical NAND output of "1" is achieved and propagated to the three fan-out branches.

In summary, we have proved through simulation that the NAND gate, fan-in and fan-out work correctly for all four input combinations for a given initial state of the nanomagnets. This is repeated in the supplementary material accompanying this paper for different initial ground states – (0, 0), (0, 1) and (1, 0) – in order to be exhaustive.

**IV Energy Considerations: Power dissipated in the magnets and the clock**

There are two important sources of energy dissipation: (i) internal energy dissipated in the magnets due to Gilbert damping and (ii) energy dissipated in the clock while charging the capacitance of the PZT layer that can be modeled as a parallel-plate capacitor.

**(i)     Internal energy dissipation**

The internal energy $E_d$ dissipated in the magnets during magnetization rotation under stress was estimated using Equation (12). The energy dissipated in the 4 nanomagnets (magnets 3, 4, 5 and 8) that comprise the NAND gate over one clock cycle varied depending on the operation performed. For example, when both inputs were "1" (Fig 2(a)) the energy dissipated was 767 kT while for both inputs set to "0" (Fig 2(b)), the energy dissipated was 948 kT. On the average, the energy dissipated in these four nanomagnets over one clock cycle is ~1000 kT. When all 12 nanomagnets are considered, the average energy dissipated over one clock cycle is ~3000 kT. Ultimately, this energy (~250kT/nanomagnet/bit) is well over the Landauer limit of kTln(2) [29] but considerably less than that dissipated in a transistor, or a nanomagnet switched with spin transfer torque, or domain wall motion or current-generated magnetic field when the gate operation is completed in 2 ns. The dissipation is governed by the (i) strength of dipole coupling needed to ensure the nanomagnets switch to the correct state with low dynamic error [30] under thermal noise and (ii) the large stress anisotropy needed to ensure that the switching is accomplished in ~ 2 ns.

**(ii)   Energy dissipated in the clock**

The energy dissipated in the clock is governed by the electrostatic potential that needs to be applied across the PZT layer to generate the stress (10 MPa) that can beat the shape anisotropy to flip magnetization and complete the gate operation in 2 ns. In this paper, we assume that the PZT layer is 40 nm thick (to ensure it is stiff compared to the magnetostrictive layer, ensuring most of the strain in it is transferred). On application of an electrostatic potential of ~50 mV across the PZT layer, an electric field of 1.25 MV/m is generated. Since the $d_{31}$ coefficient of PZT is ~ $-10^{-10}$ m/V [27], a strain of ~$1250 \times 10^{-6}$ is generated and transferred to the Terfenol-D layer resulting in a stress ~10 MPa since the Young's modulus of Terfenol-D is $8 \times 10^{10}$ Pa.

Next, we estimate the capacitance of the ~40 nm thick PZT layer of surface area 101.75 nm × 98.25 nm and thickness 40 nm as 1.74 fF (relative dielectric constant ~ 1000 [27]). Thus the energy dissipated in applying 50 mV, then switching it to -50 mV and discharging to zero (to generate the stress cycle shown in Fig 1 c) is ~ 3200 kT per nanomagnet if this is done abruptly (the energy dissipated in charging the capacitor abruptly with a square wave pulse is $\frac{1}{2}CV^2$ so that charging it up to +V from 0, reversing it to −V, and then discharging it back to 0 dissipates an energy of $3CV^2$. In contrast, driving an

$$E_{clock} = \pi \frac{V^2}{R} \omega (RC)^2$$

LCR circuit with a sinusoidal source dissipates, *resulting in an energy saving by a factor* $\frac{\pi \omega RC}{3}$.

Abrupt (non-adiabatic) switching with a square wave pulse will cause a total energy dissipation of ~ 40,000 kT in the clocking circuit (or more than 10 times the internal energy dissipated in the nanomagnets). However, if the LCR circuit is driven with a sinusoidal voltage of low frequency (large time period), then clocking becomes quasi-adiabatic. This reduces dissipation considerably because of the large energy saving factor. From Equation 10 (b), we can see that if the time period is much larger than RC, the dissipation is greatly reduced. In our case, we assume that the PZT layer is electrically accessed with a silver wire of resistivity 2.6 μΩ-cm [31] so that an access line of length 10 μm and cross section 50 nm×50 nm has resistance ~100 Ω. Hence, the *RC* time constant is ~0.174 ps. The clock period is 2 ns, so that the reduction factor $\frac{\pi \omega RC}{3} = 5.47 \times 10^{-4}$. This makes the dissipation in the clock only about 22 kT, which is negligible compared to the internal energy dissipation of 3000 kT.

**V Conclusions**

We have modeled the nonlinear magnetization dynamics of an all-multiferroic nanomagnetic NAND gate with fan-in/fan-out and shown that a throughput of ~ 1 bit per 2 ns and latency ~4 ns can be achieved, so that the clock rate can be 0.5 GHz. Such a gate circuit is estimated to dissipate ~ 3000 kT/clock cycle internally in the 12 nanomagnets combined and much less energy (20 kT/clock cycle) in the external access circuitry for the clock signal, if we use a 4-phase clocking scheme with a sinusoidal voltage source driving an LCR circuit.

All this begs the question as to whether it is possible to reduce the internal energy dissipation by some appropriate scheme. This was discussed in ref. [2]. Imagine a magnet made of a material that has no Gilbert damping $(\alpha = 0)$. If we remove the shape anisotropy barrier and make the magnet isotropic (circular disk), then a magnetic field applied perpendicular to the magnet's plane will make the magnetization vector precess around it without any damping in accordance with the first term in the right hand side of Equation (3). There is now no internal dissipation. The magnetic field however must be removed *precisely* at the juncture when the magnetization completes $180^0$ rotation if we wish to flip the bit. This requires exact precision; otherwise, the magnet will either not have completed $180^0$ rotation, or overshot, resulting in more than $180^0$ rotation. This error will continue to build up with time and finally become too large to endure. In other words, there is no fault tolerance. This is a well-known problem that has been discussed by numerous authors starting from the Fredkin billiard ball computer which can compute without dissipating energy [32], but cannot tolerate any error. Clearly, if we require fault tolerance, we must have damping, and hence some internal dissipation. In the presence of damping, fluctuations can deviate the magnetization from the desired orientation (minimum energy state), but the latter will return to the correct orientation (minimum energy state) by dissipating energy. Therefore, the dissipation in the clocking circuit can be eliminated by adiabatic approaches (increasingly slow switching), but the internal dissipation must remain for the sake of fault tolerance.

The internal energy dissipated in the magnet must be provided by the power source driving the clock. This source need not dissipate any energy to raise and lower the barrier separating the logic bits as long as we raise and lower the barrier adiabatically, but it must dissipate some energy internally in the logic device to maintain fault tolerance.

**Acknowledgement:** This work is supported by the US National Science Foundation under the "Nanoelectronics Beyond the Year 2020" grant 1124714.

Table.1 Material parameters and geometric design for Terfenol-D

| | |
|---|---|
| $(3/2)\lambda_s$ | $9\times10^{-4}$ [23], [24] |
| $M_s$ | $0.8\times10^6$ A m$^{-1}$ |
| Young's modulus | $8\times10^{10}$ Pa [25] |
| $\alpha$ | 0.1 [26] |
| **Dimension** $a\times b\times t$ | $101.75nm\times 98.25nm\times 10nm$ |
| $r$ | 200 nm |

**Figure captions**

**Fig 1(a):** Design of all multiferroic NAND gate with input "logic wires" and fan-out. The magnetization directions shown depict the correct initial (ground) state corresponsing to input-1 = 1 and input-2 = 1.

**Fig 1 (b):** Two nanomagnets whose hard axes are at an angle γ to the line joining their centers .

**Fig 1 (c):** 4 phase clock showing sinusoidal stress applied to the nanomagnets.

**Fig 2:** LLG simulation of magnetization dynamics of all magnets in the chain with initial (ground) states corresponding to the input-1=1 and input-2=1. Thereafter, inputs are changed to:
 **(a)** Input-1=1 and input-2=1 followed by applying 4-phase clock, which results in output=0.
 **(b)** Input-1=0 and input-2=0 followed by applying 4-phase clock, which results in output=1.
 **(c)** Input-1=0 and input-2=1 followed by applying 4-phase clock, which results in output=1.
 **(d)** Input-1=1 and input-2=0 followed by applying 4-phase clock, which results in output=1.

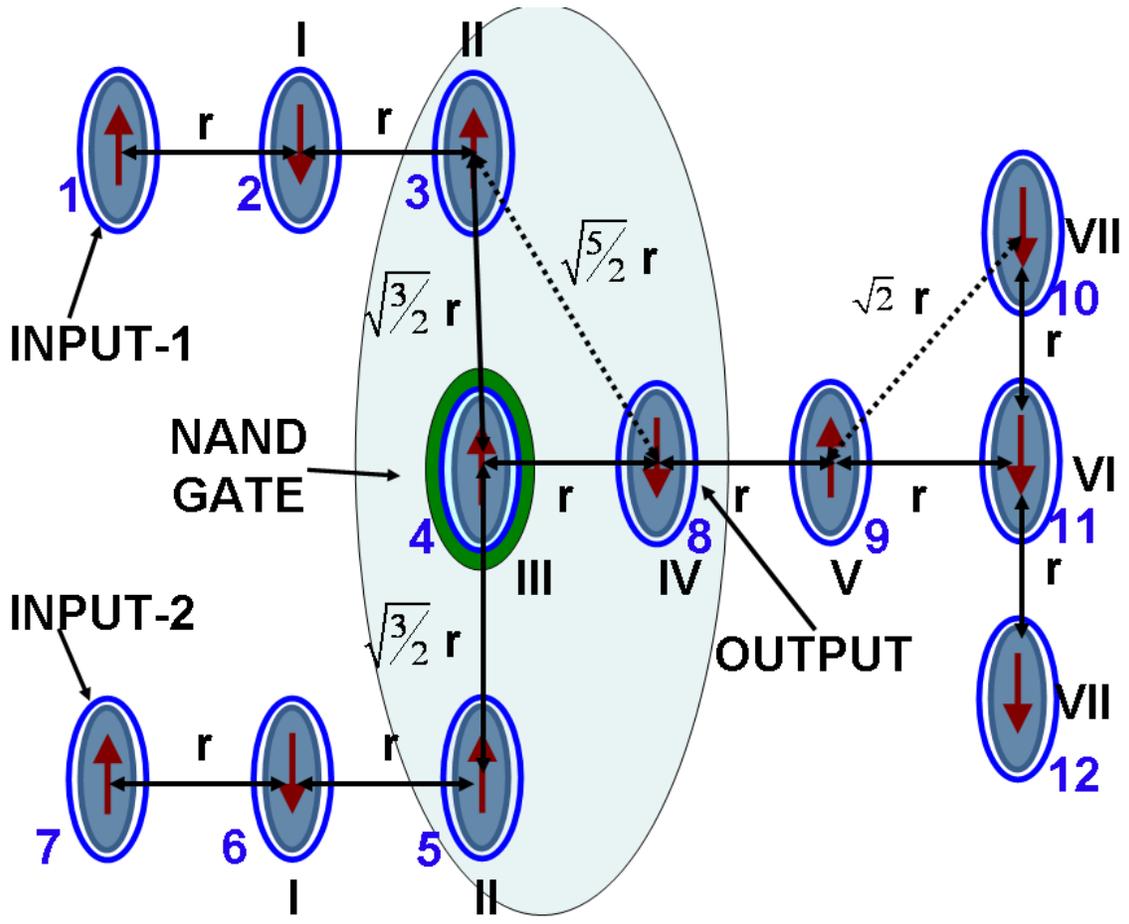

**Fig 1(a)**

**Fig 1 (b)**

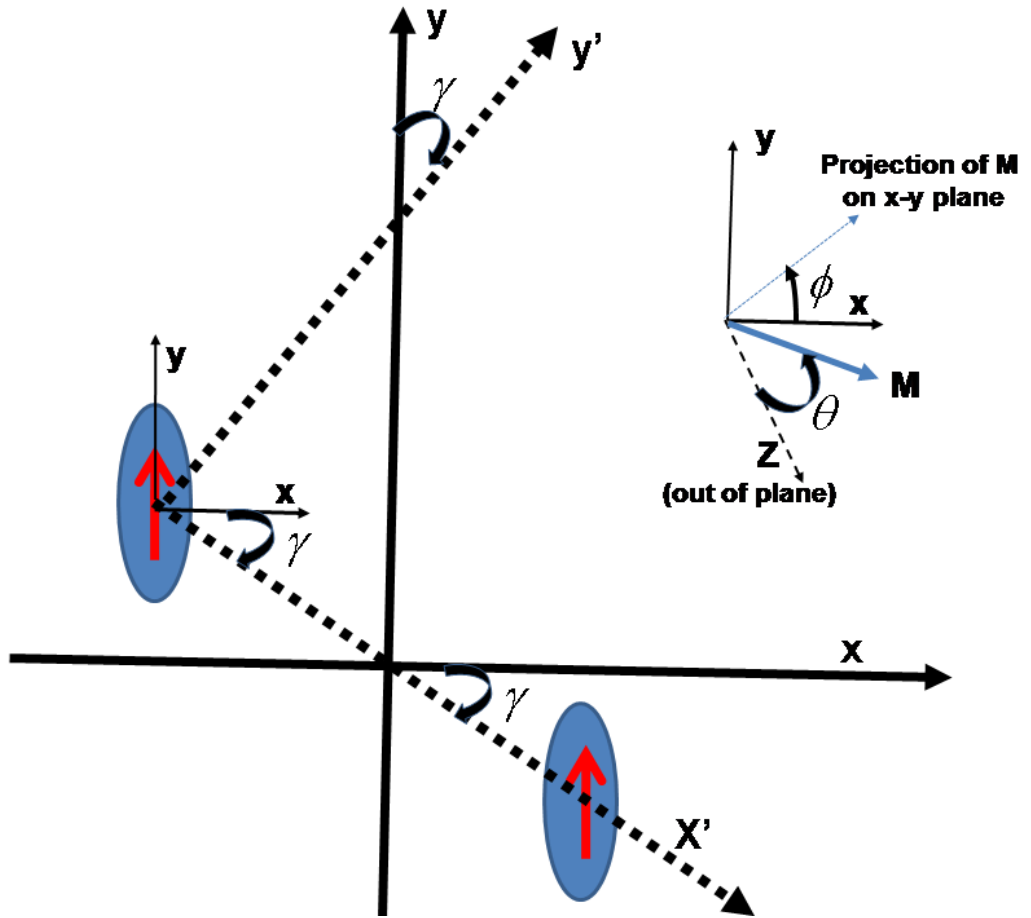

**Fig 1 (c): 4 phase clock**

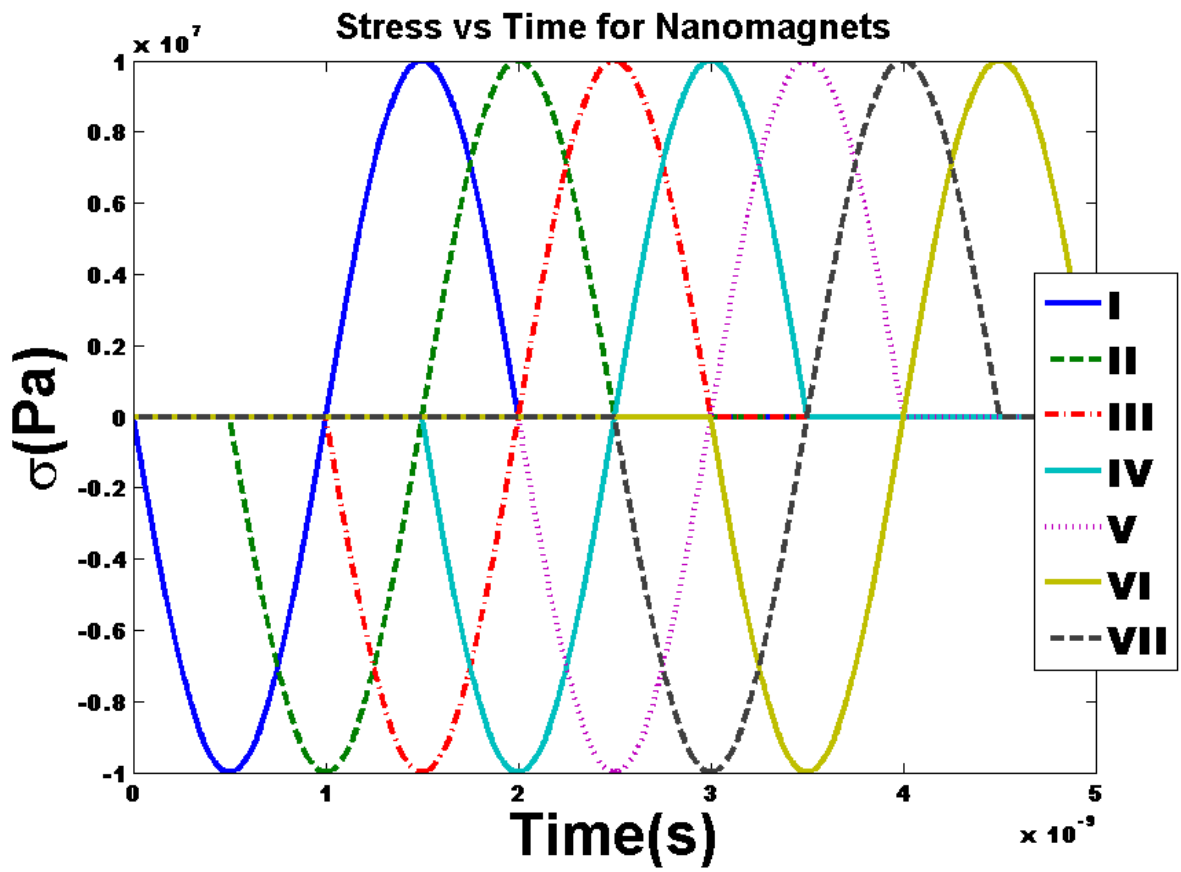

**Fig.2 (a)**

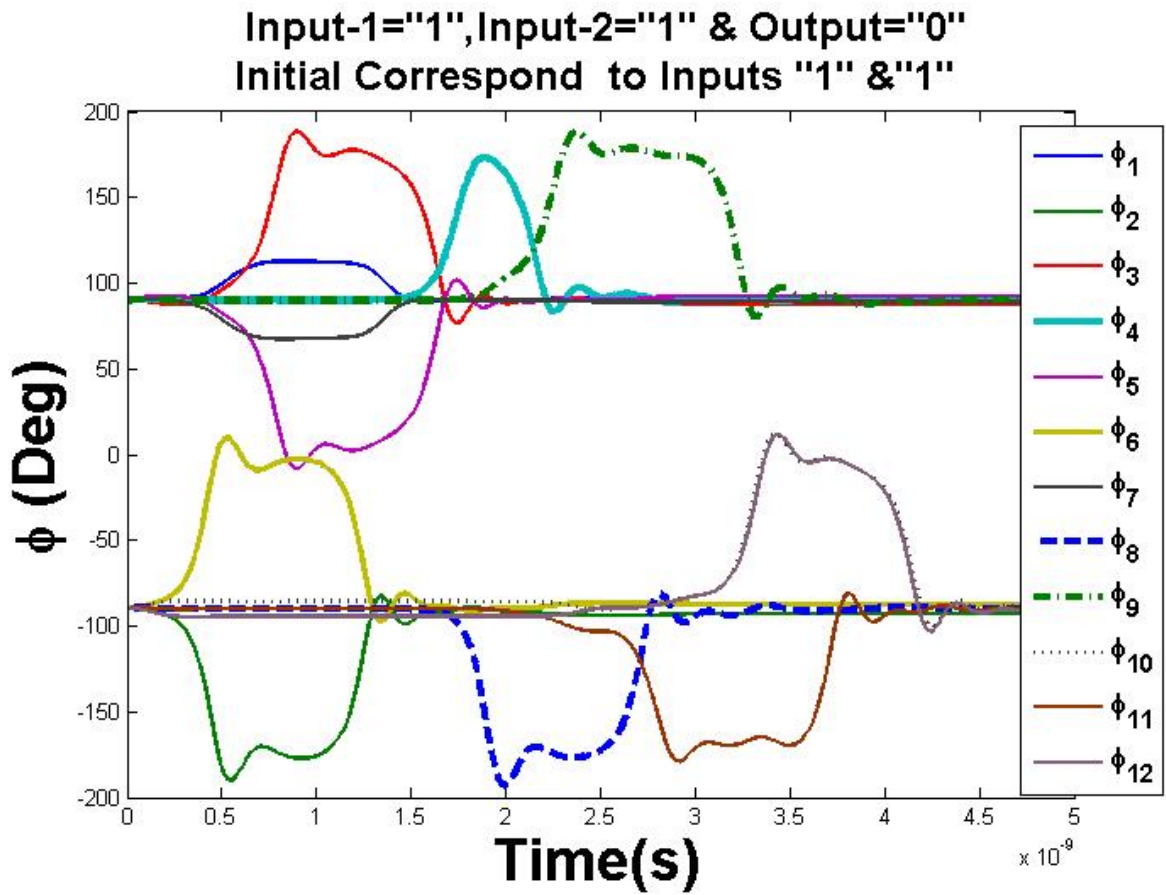

**(b)**

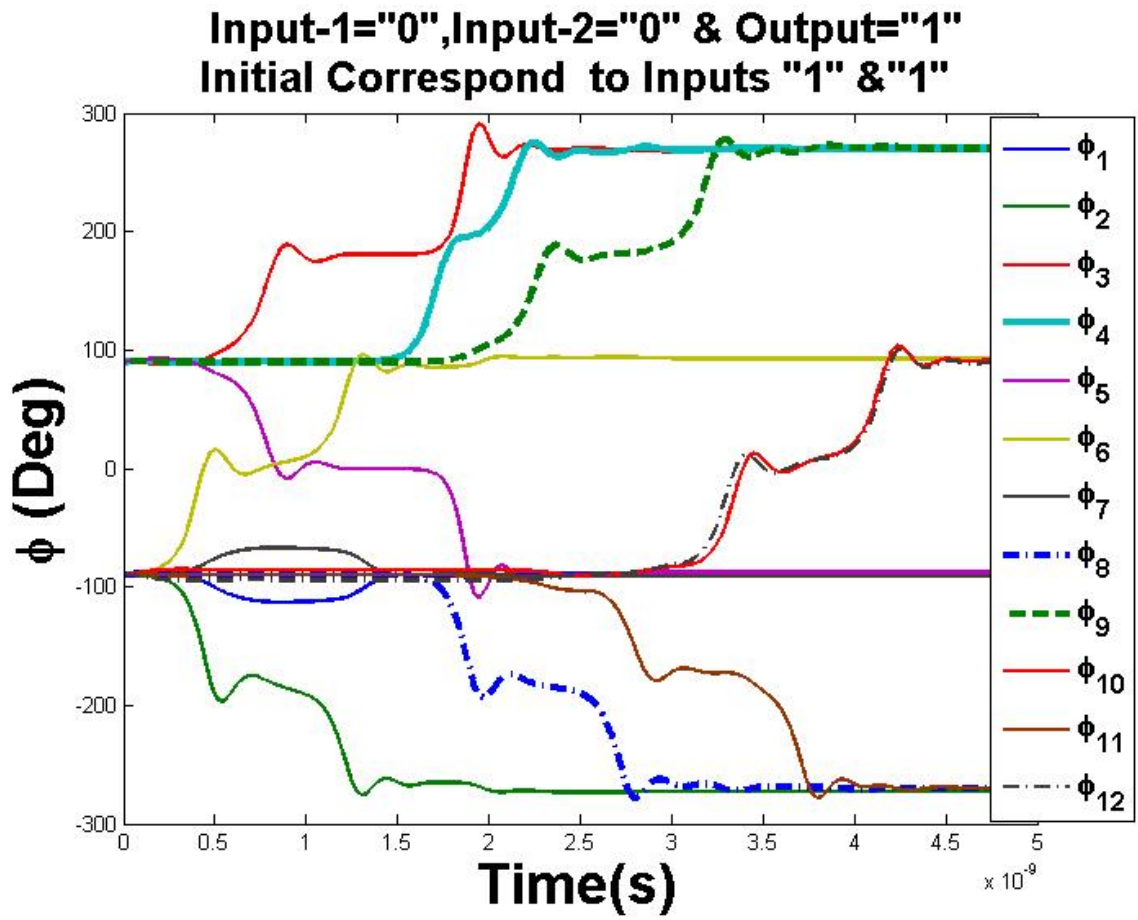

**(c)**

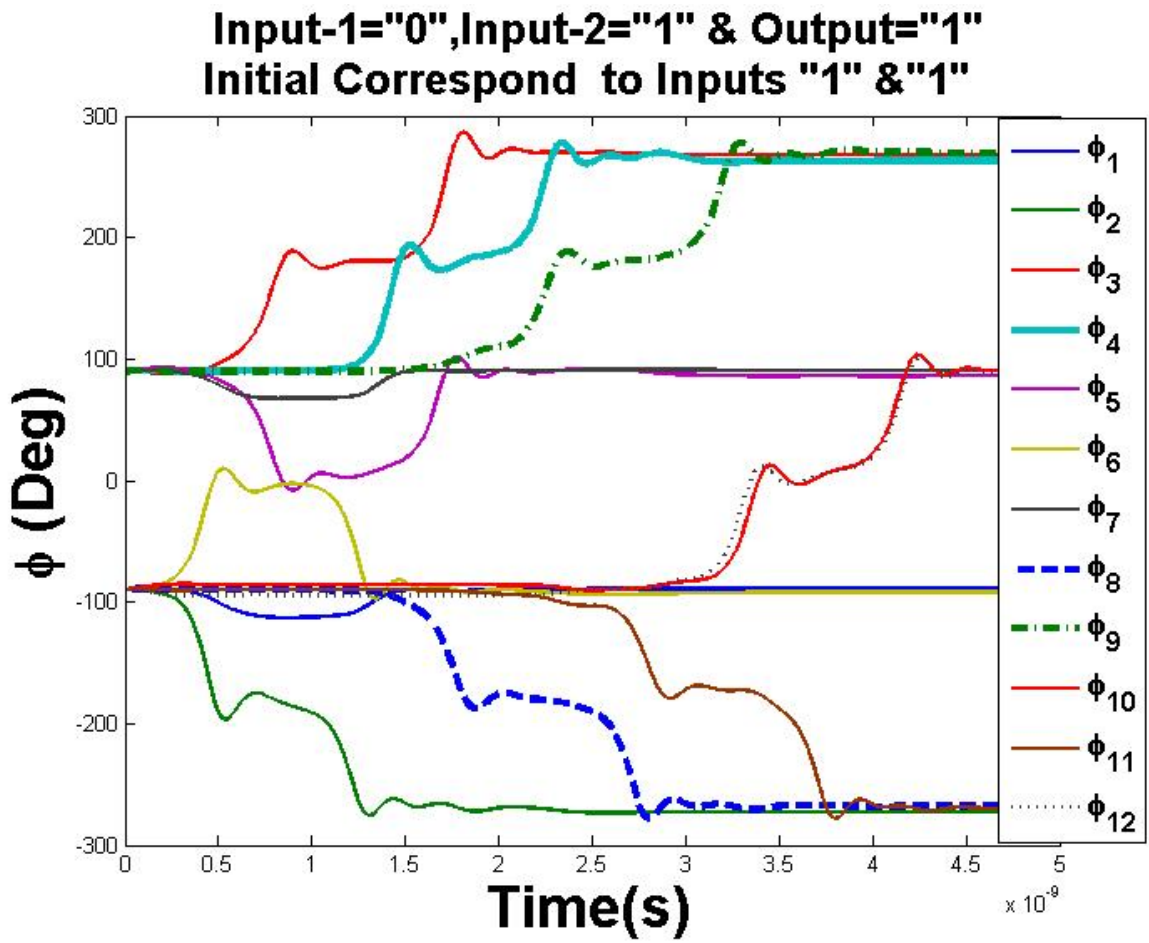

**(d)**

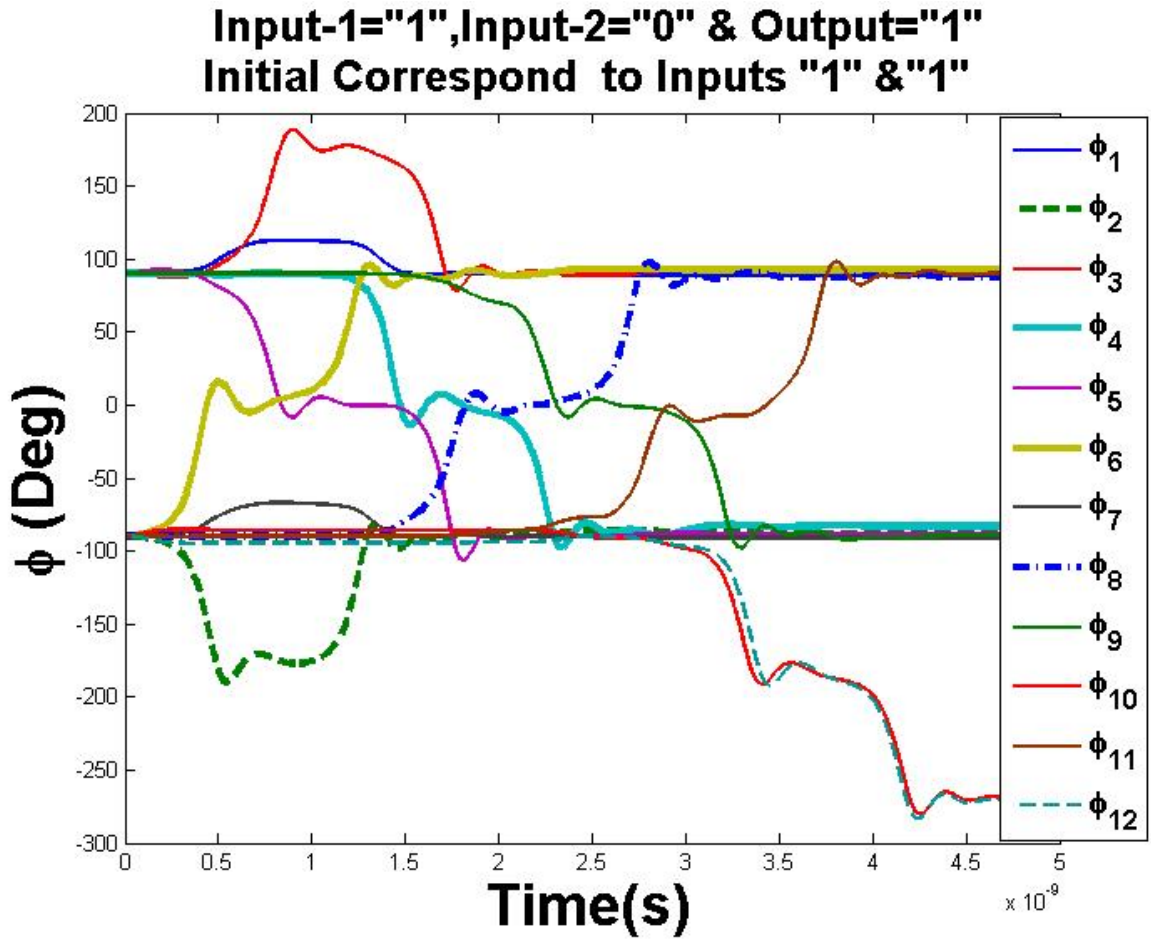

# Magnetization Dynamics, Throughput and Energy Dissipation in a Universal Multiferroic Nanomagnetic Logic Gate

Mohammad Salehi-Fashami[a], Jayasimha Atulasimha[a], Supriyo Bandyopadhyay[b]
Email: {salehifasham, jatulasimha, sbandy}@vcu.edu

(a) Department of Mechanical Engineering,

(b) Department of Electrical and Computer Engineering,

Virginia Commonwealth University, Richmond, VA 23284, USA.

In the main paper we theoretically demonstrated the switching dynamics for the case when the ground state (initial conditon) of all nanomagnets in the input logic wires, NAND gate, and fan-out correspond to the correct states for input-1=1 and input-2=1. Thereafter, different inputs combinations as shown in Table-1 were applied and it was theoretically demonstrated that the magnetizations switch (from the above ground state) to give the correct output states. Thus, we proved that this device works infallibly for all combinations of inputs when the initial (ground) state of the nanomagnetic logic wires, NAND gate, and fan-out correspond to the correct states for input-1=1 and input-2=1. (NOTE: That Case 1 is trivial as the ground state already corresponds to the correct output, but it is still important to verify that the gate works correctly for this state.)

**Table 1**. Different initial conditions.

| Different Inputs to NAND gate | INPUT-1 | INPUT-2 | OUTPUT |
|---|---|---|---|
| 1. | 1 | 1 | 0 |
| 2. | 0 | 0 | 1 |
| 3. | 0 | 1 | 1 |
| 4. | 1 | 0 | 1 |

In the supplement we present 12 more switching diagrams to show that the NAND gate works for 3 other initial conditions of the nanomagnetic logic wires, NAND gate, and fan-out:

**CASE I. Initial (groud) state corresponding to input-1=0 and input-2=0**

Fig S1: shows the correct initial (ground) state for input-1=0 and input-2=0.

Fig S2 (a)-(d):show the switching diagrams for four different combinations of input-1 and input-2. In each case the correct output corresponsing to a NAND gate is acheived.

**CASE II. Initial (groud) state corresponding to input-1=0 and input-2=1**

Fig S3: shows correct initial (ground) state for input-1=0 and input-2=1.

Figures S4 (a)-(d):show the switching diagrams for four different combinations of input-1 and input-2. In each case the correct output corresponsing to a NAND gate is acheived.

**CASE III. Initial (groud) state corresponding to input-1=1 and input-2=0**

Fig S5: shows correct initial (ground) state for input-1=1 and input-2=0.

Figures S6 (a)-(d):show the switching diagrams for four different combinations of input-1 and input-2. In each case the correct output corresponsing to a NAND gate is acheived.

These simulations in the supplemant, taken together with the simulations in the main paper, prove that the NAND gate works for different combinations of inputs for all possible initial conditions.

**Figure S1:** Shows the correct initial (ground) state of the nanomagnets corresponding to input-1=0 and input-2=0.

**Figure S2:** LLG simulation of magnetization dynamics of all magnets in the chain with initial (ground) states corresponding to the input-1=0 and input-2=0. Thereafter, inputs are changed to:

(a) Input-1=1 and input-2=1 followed by applying 4-phase clock, which results in output=0.

(b) Input-1=0 and input-2=0 followed by applying 4-phase clock, which results in output=1.

(c) Input-1=0 and input-2=1 followed by applying 4-phase clock, which results in output=1.

(d) Input-1=1 and input-2=0 followed by applying 4-phase clock, which results in output=1.

**Figure S3:** Shows the correct initial (ground) state of the nanomagnets corresponding to input-1=0 and input-2=1.

**Figure S4:** LLG simulation of magnetization dynamics of all magnets in the chain with correct initial (ground) state corresponding to the input-1=0 and input-2=1. Thereafter, inputs are changed to:

(a) Input-1=1 and input-2=1 followed by applying 4-phase clock, which results in output=0.

(b) Input-1=0 and input-2=0 followed by applying 4-phase clock, which results in output=1.

(c) Input-1=0 and input-2=1 followed by applying 4-phase clock, which results in output=1.

(d) Input-1=1 and input-2=0 followed by applying 4-phase clock, which results in output=1.

**Figure S5:** Shows the correct initial (ground) state of the nanomagnets corresponding to input-1=1 and input-2=0.

**Figure S6:** LLG simulation of magnetization dynamics of all magnets in the chain with correct initial (ground) state corresponding to the input-1=1 and input-2=0. Thereafter, inputs are changed to:

(a) Input-1=1 and input-2=1 followed by applying 4-phase clock, which results in output=0.

(b) Input-1=0 and input-2=0 followed by applying 4-phase clock, which results in output=1.

(c) Input-1=0 and input-2=1 followed by applying 4-phase clock, which results in output=1.

(d) Input-1=1 and input-2=0 followed by applying 4-phase clock, which results in output=1.

**Supplementary Figures :**

**Case I.**

**Fig.S1**

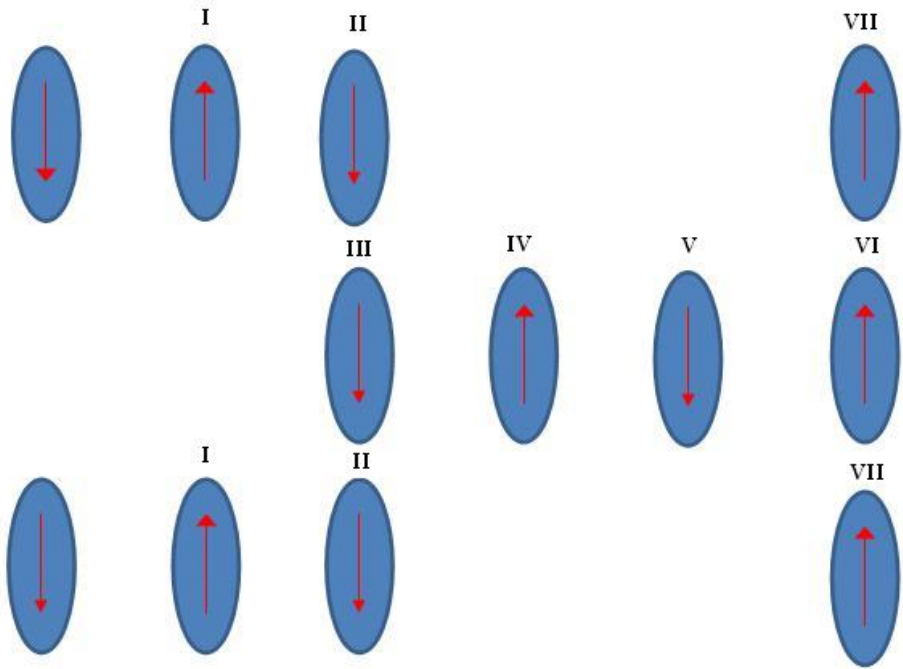

**Fig.S2(a)**

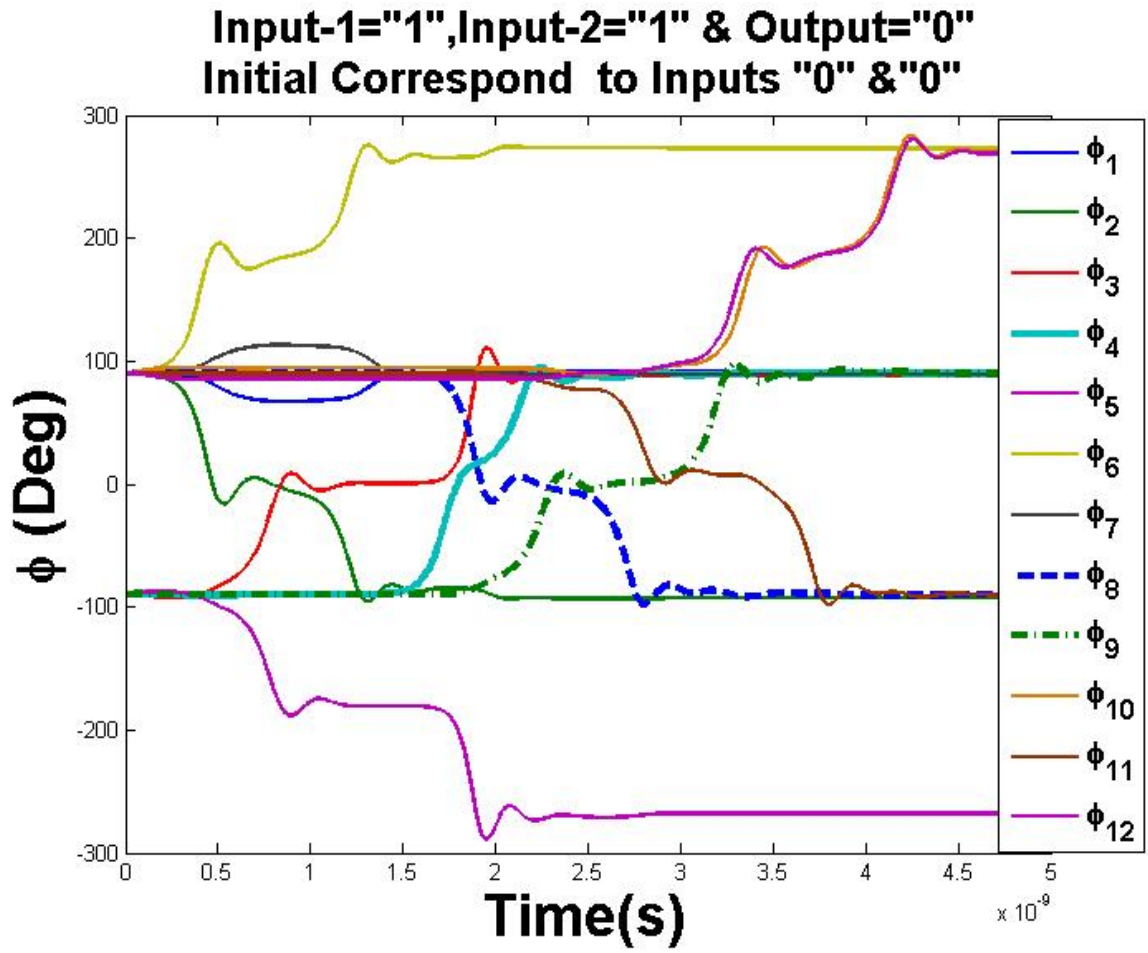

**Fig.S2(b)**

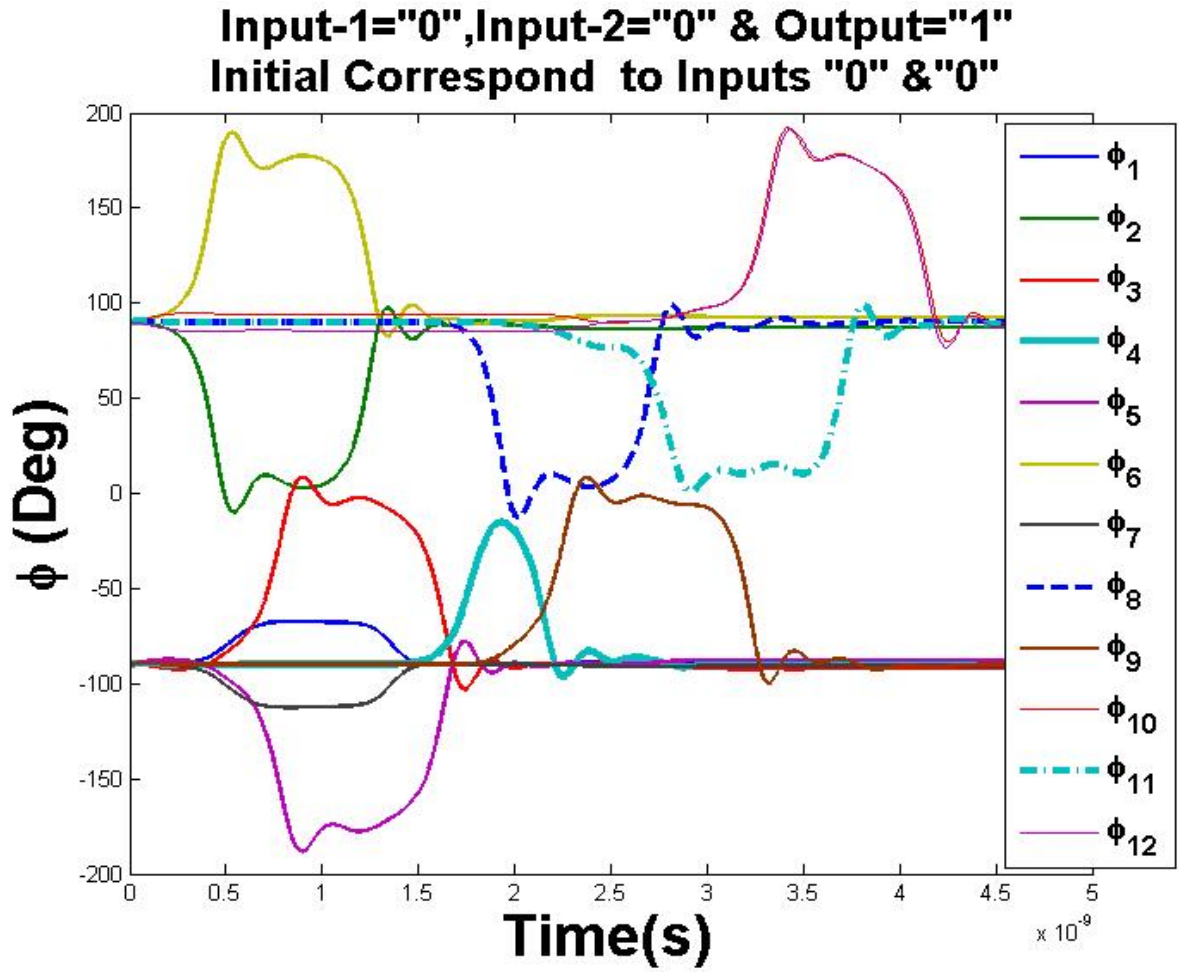

**Fig.S2(c).**

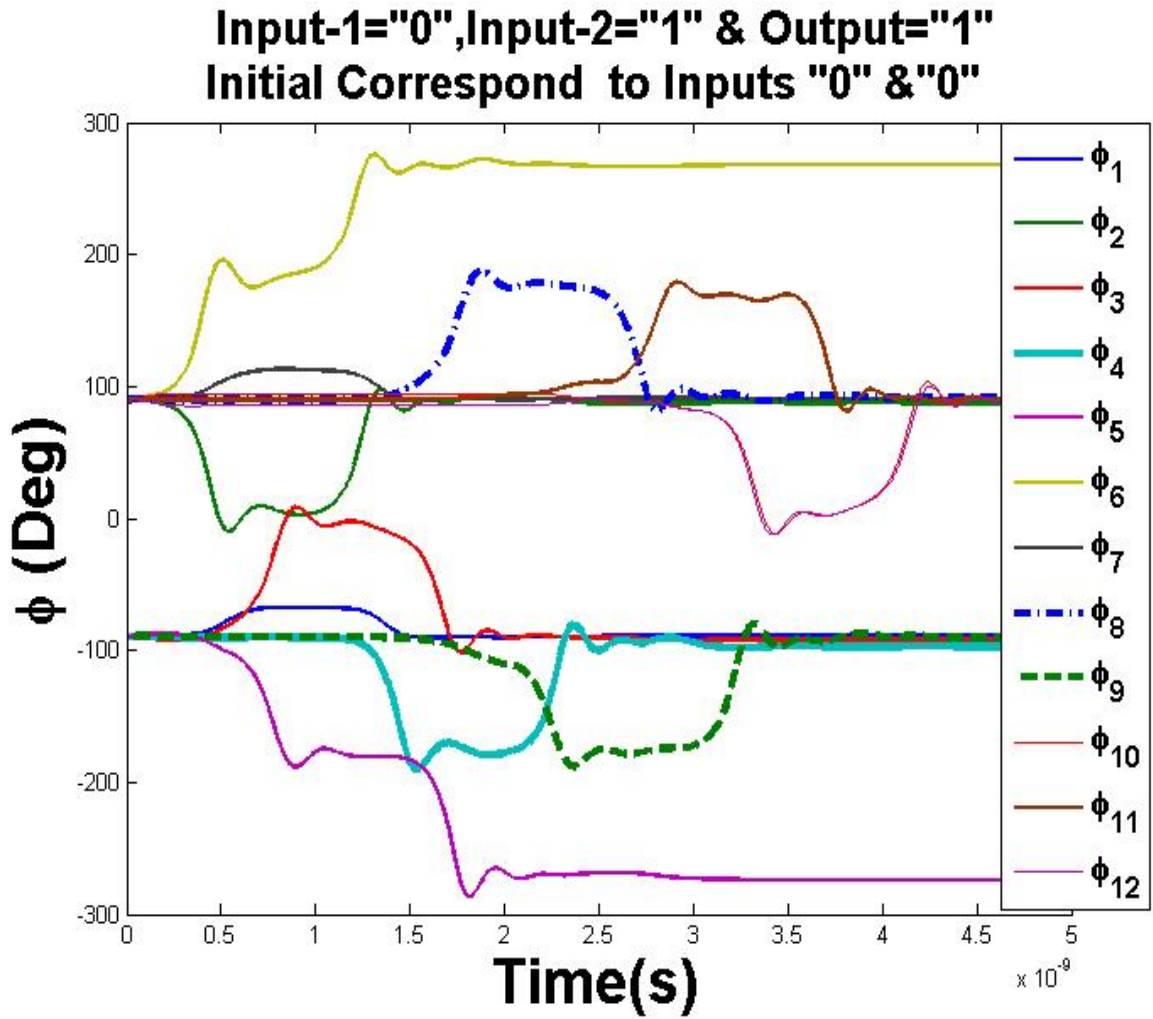

**Fig.S2(d)**

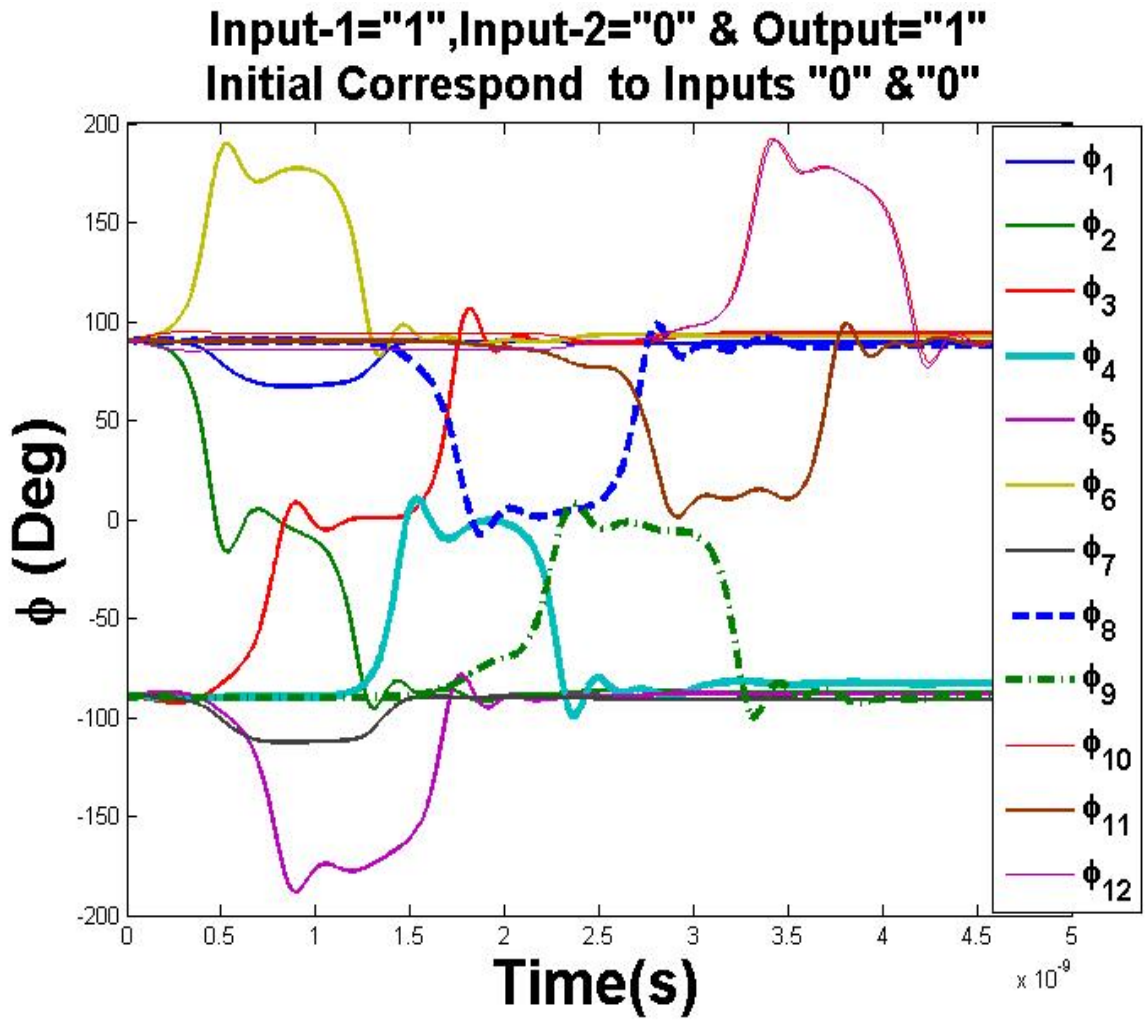

**Case II**

**Fig.S3.**

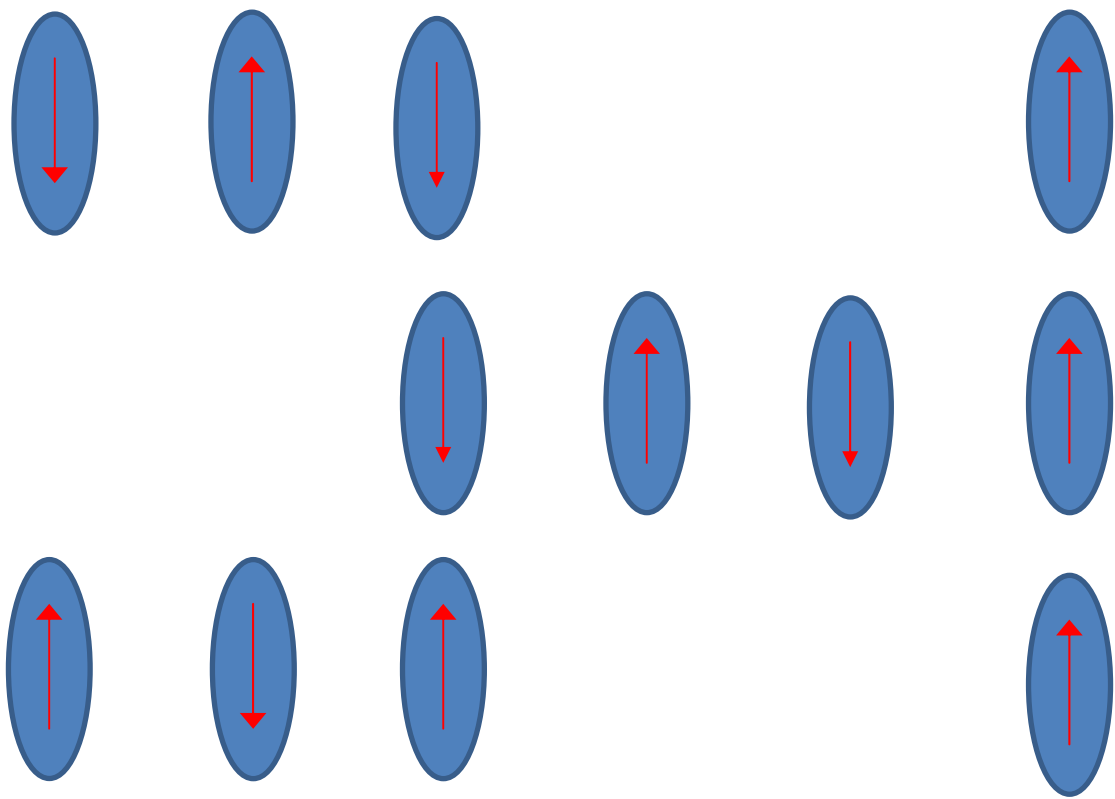

**Fig.S4(a)**

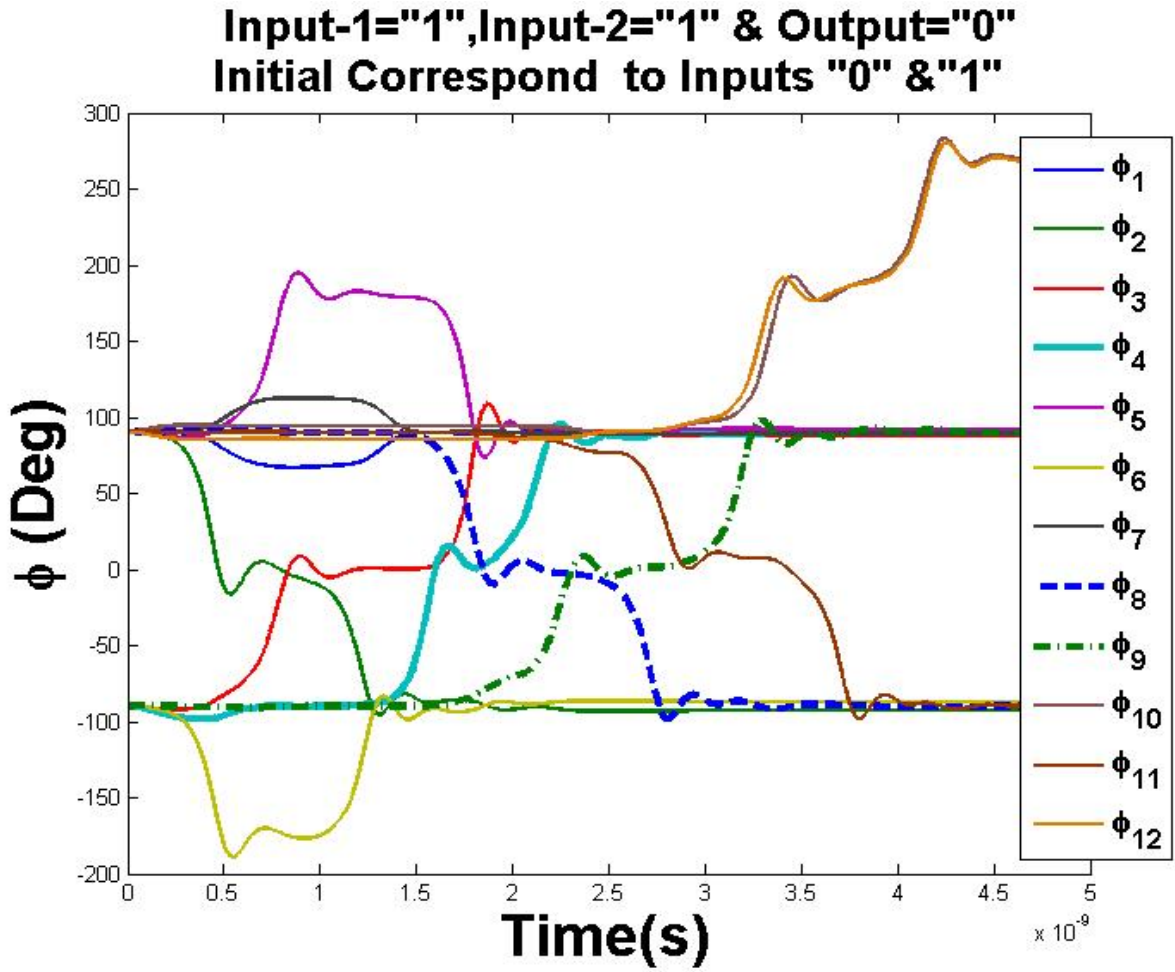

**Fig.S4(b).**

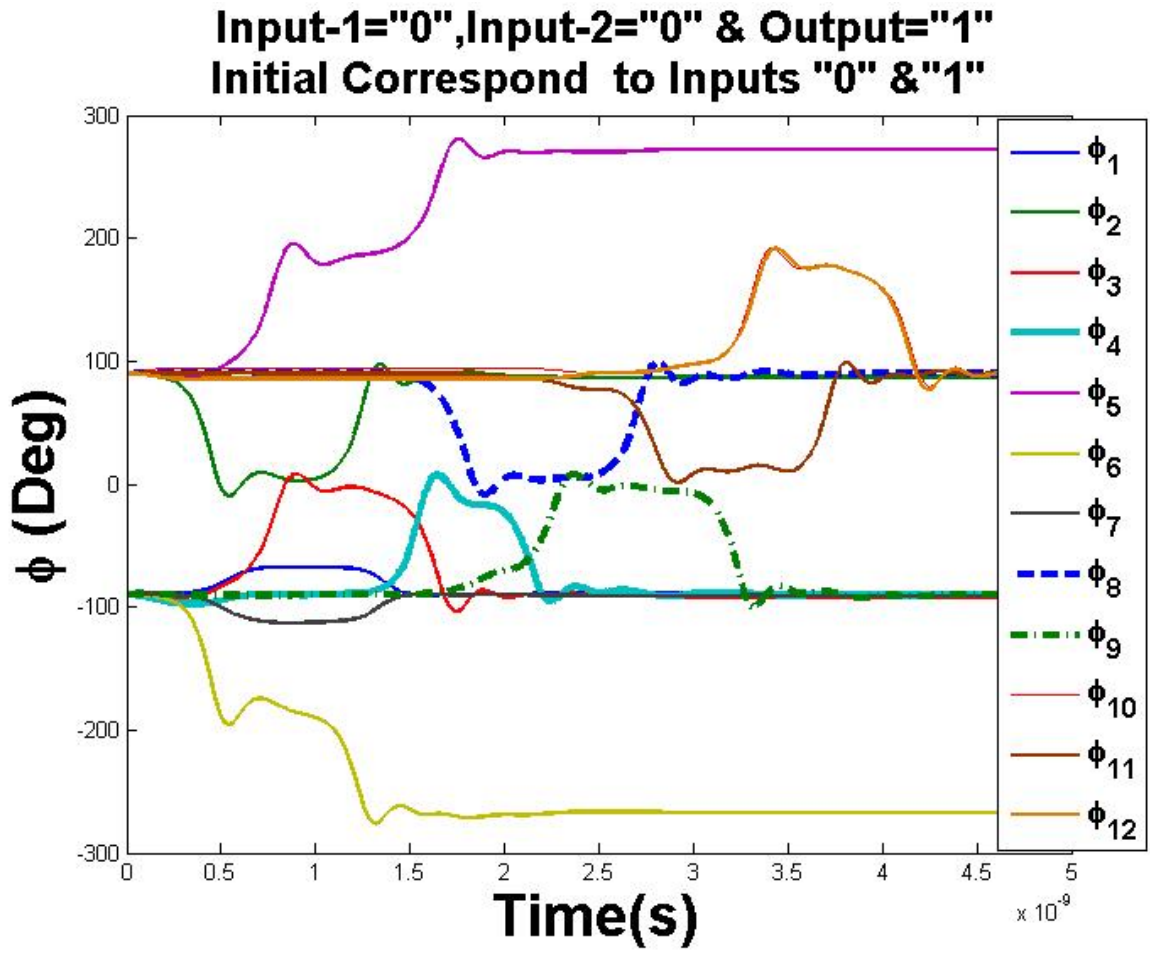

**Fig.S4(c).**

Figure: Input-1="0", Input-2="1" & Output="1" Initial Correspond to Inputs "0" & "1". Plot of $\phi$ (Deg) vs Time(s), showing traces $\phi_1$ through $\phi_{12}$.

**Fig.S4(d).**

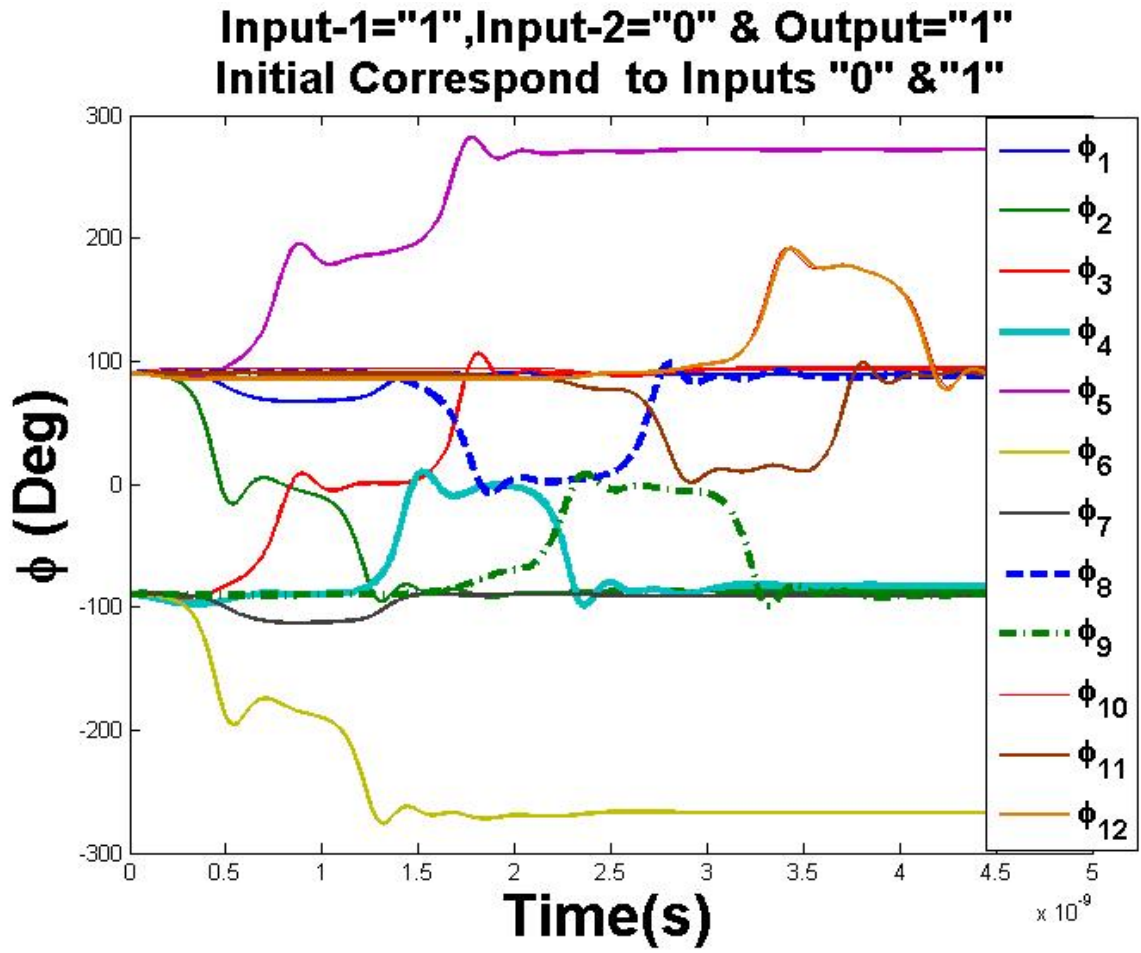

**Case III:**

**Fig.S5.**

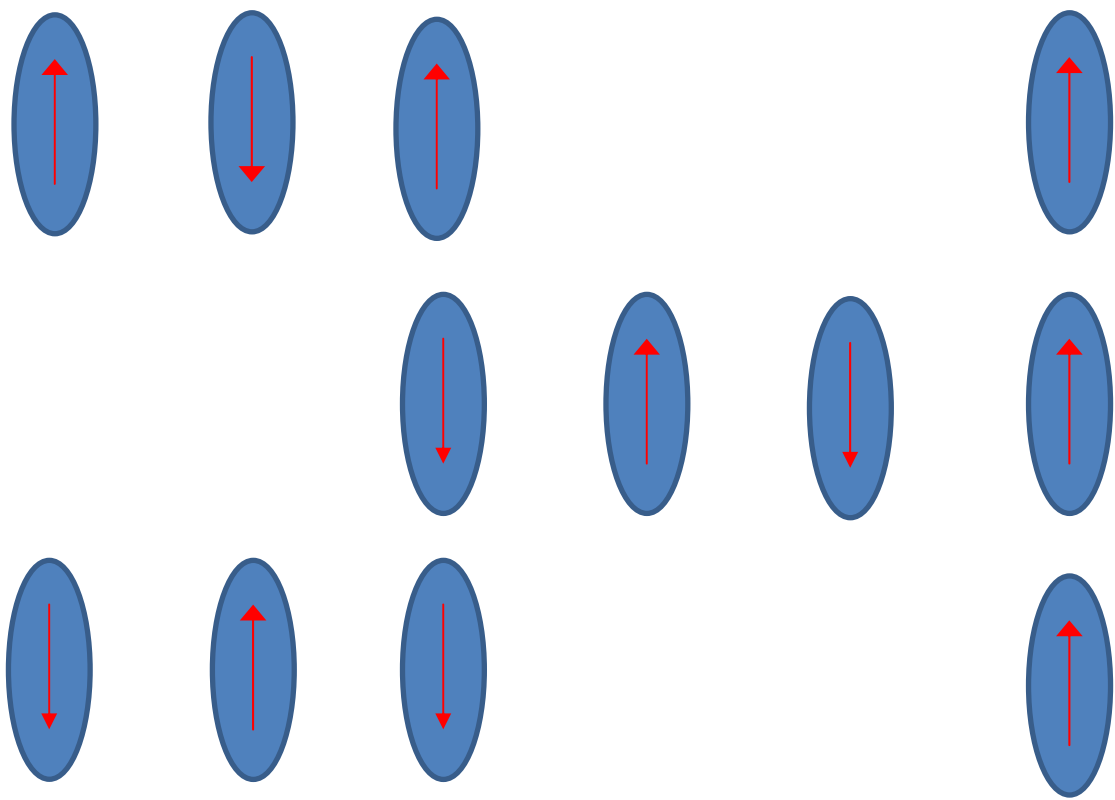

**Fig.S6(a)**

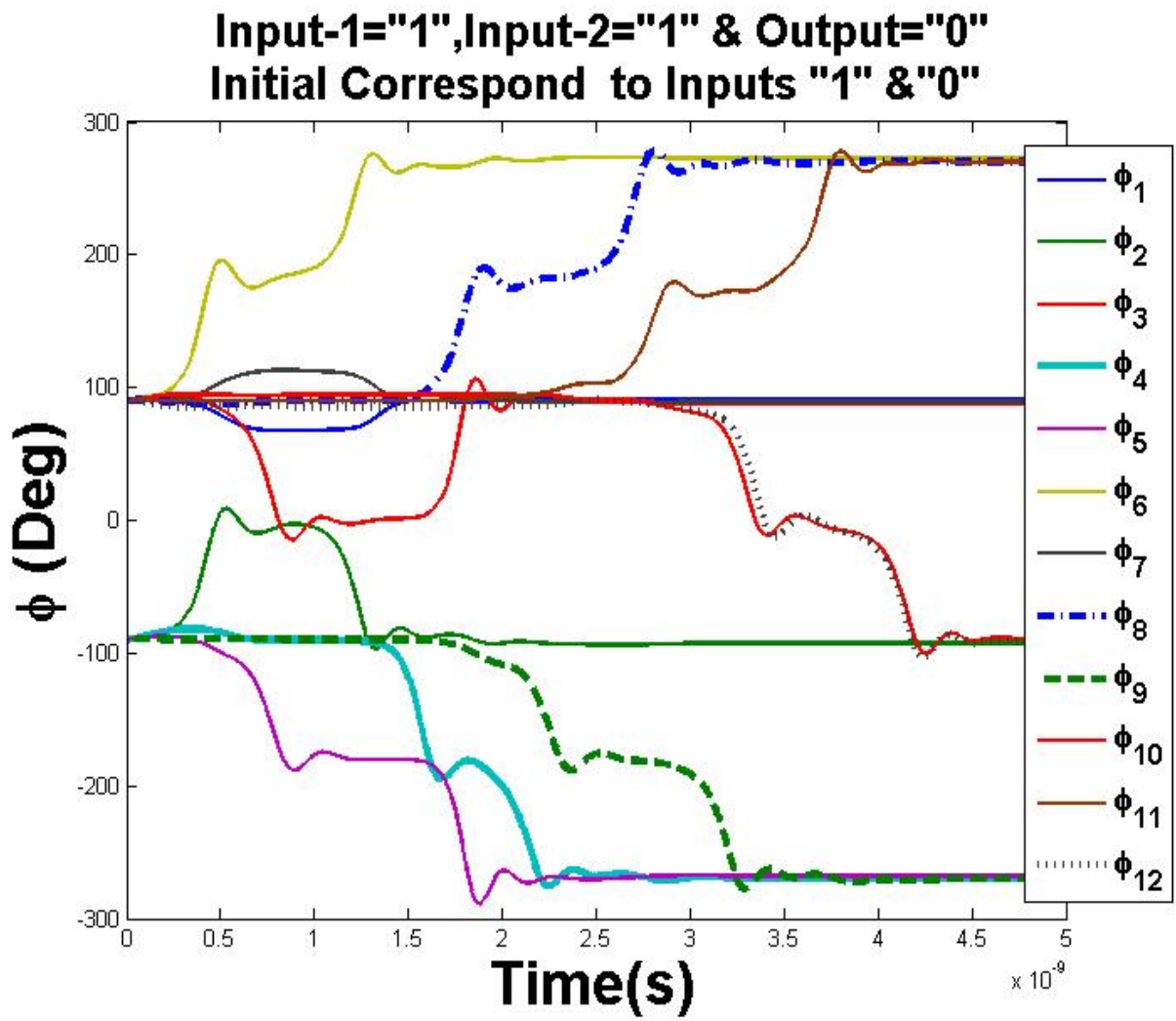

**Fig.S6(b)**

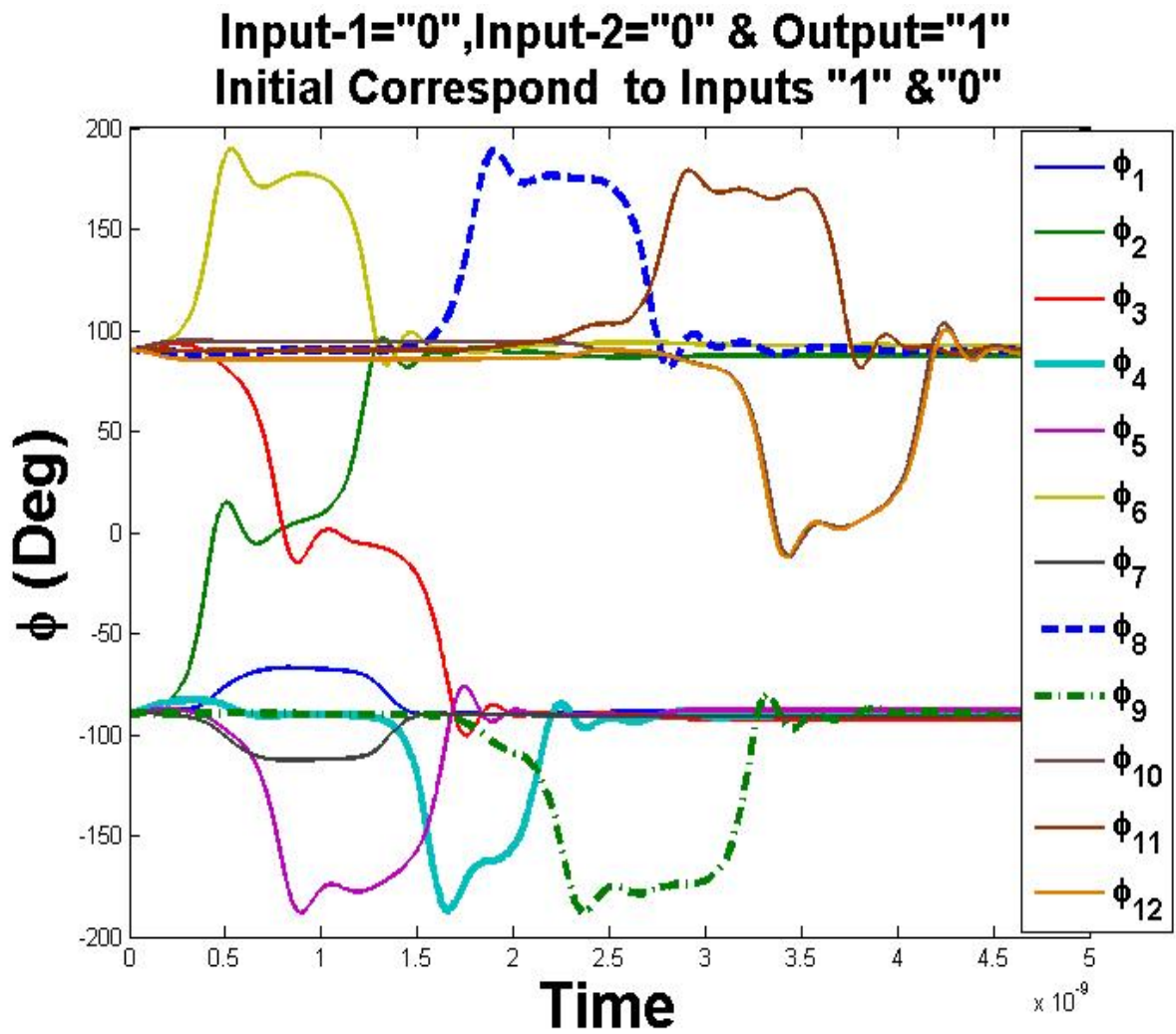

**Fig.S6(c)**

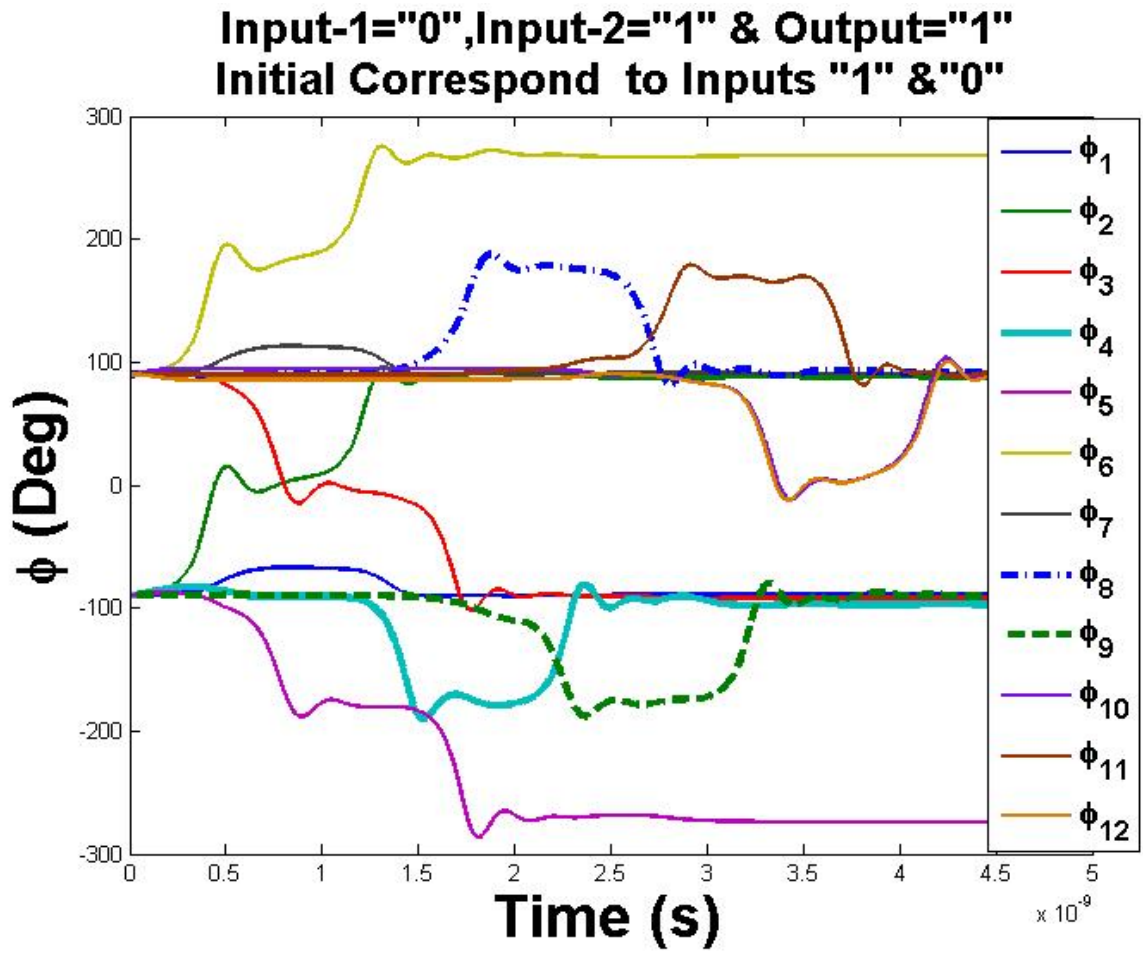

**Fig.S6(d)**

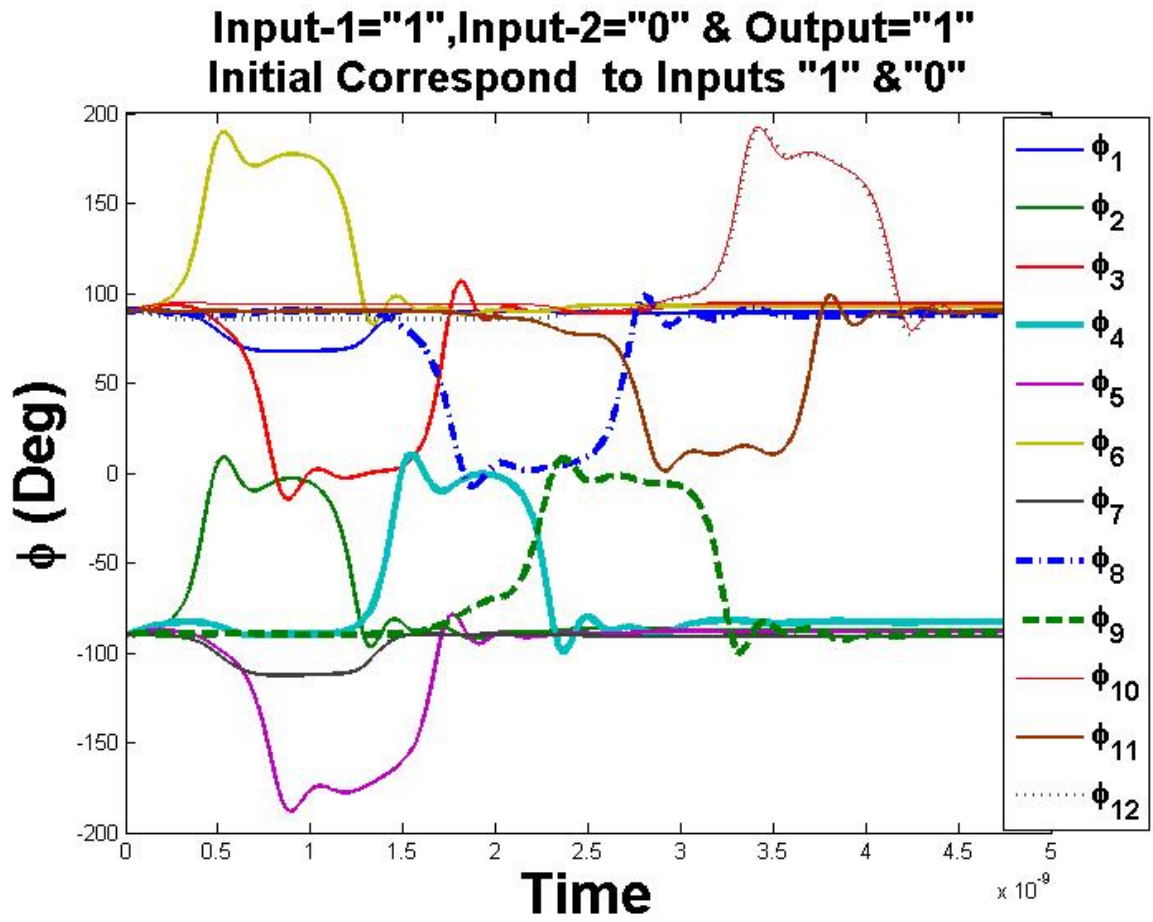